\documentclass[fleqn,usenatbib]{mnras}

\usepackage{graphicx}
\usepackage{amsmath,amssymb}
\usepackage{pdflscape}
\usepackage{newtxtext,newtxmath}

\usepackage[T1]{fontenc}
\usepackage{ulem}

\DeclareRobustCommand{\VAN}[3]{#2}
\let\VANthebibliography\thebibliography
\def\thebibliography{\DeclareRobustCommand{\VAN}[3]{##3}\VANthebibliography}

\title[Protoclusters at $z\sim3$]{Diverse Protoclusters at $\mathbf{z\sim3}$ Disclosed by Systematic Spectroscopy}

\author[J. Toshikawa et al.]{Jun Toshikawa,$^1$\thanks{E-mail: toshikawa@astr.tohoku.ac.jp} % ORCID:0000-0001-5394-242X
Kentaro Koretomo,$^2$ % ORCID:0009-0001-5183-2945
Nobunari Kashikawa,$^{2,3}$ % ORCID:0000-0003-3954-4219
Stijn Wuyts,$^4$ % ORICD:0000-0003-3735-1931
Marcin Sawicki,$^5$ % ORCID:0000-0002-7712-7857
\newauthor Malcolm N. Bremer,$^6$
Yoshiaki Ono,$^7$ % ORCID:0000-0001-9011-7605
Mariko Kubo$^8$ % ORCID:0000-0002-7598-5292
and Tomoki Saito$^9$\\
$^1$Astronomical Institute, Tohoku University, 6-3, Aramaki, Aoba, Sendai, Miyagi, 980-8578, Japan\\
$^2$Department of Astronomy, University of Tokyo, 7-3-1, Hongo, Bunkyo, Tokyo 113-0033, Japan\\
$^3$Center for the Early Universe, The University of Tokyo, 7-3-1 Hongo, Bunkyo, Tokyo, 113-0033, Japan\\
$^4$Department of Physics, University of Bath, Claverton Down, Bath BA2 7AY, UK\\
$^5$Department of Astronomy \& Physics and Institute for Computational Astrophysics, Saint Mary’s University, 923 Robie Street, Halifax, NS B3H 3C3, Canada\\
$^6$School of Physics, HH Wills Physics Laboratory, University of Bristol, Tyndall Avenue, Bristol BS8 1TL, UK\\
$^7$Institute for Cosmic Ray Research, The University of Tokyo, 5-1-5 Kashiwa-no-Ha, Kashiwa, Chiba, 277-8582, Japan\\
$^8$Department of Physics and Astronomy, School of Science, Kwansei Gakuin University, Gakuen Uegahara, Sanda, Hyogo 669-1330, Japan\\
$^9$National Institute of Information and Communication Technologies, 893-1 Hirai, Kashima, Ibaraki 314-8501, Japan
}

\date{Accepted XXX. Received YYY; in original form ZZZ}

\pubyear{2026}

\begin{document}
\label{firstpage}
\pagerange{\pageref{firstpage}--\pageref{lastpage}}
\maketitle

\begin{abstract}
We present the results from follow-up spectroscopy of three protocluster candidates at $z\sim3$, selected as overdense regions of $U$-dropout galaxies in the wide-field imaging of the Hyper SuprimeCam Subaru Strategic Programme. All three targets exhibit tight clustering in three-dimensional space, i.e. on the sky ($>4\sigma$ overdensity) and in redshift ($>98.4\%$ confidence), making for clear evidence of genuine protoclusters. Since these three protoclusters are identified by the same dataset and technique, they represent a good sample to study the diversity of protocluster properties with minimal observational biases. Although the average population across the three protoclusters is consistent with field galaxies, each protocluster individually exhibits distinguishing properties from each other or from field galaxies. In one protocluster, member galaxies stand out by a higher Ly$\alpha$ equivalent width ($EW_0$), and massive members that are still highly star-forming. In contrast, relatively more suppressed star formation appears in another protocluster where the Ly$\alpha$ $EW_0$ of member galaxies is systematically lower. These variations can be attributed to differences in the developmental phase of protoclusters. If the presence of quenching galaxies indicates a more mature system, galaxies with stronger Ly$\alpha$ emissions may be more prevalent in the earlier stages of cluster formation. The uniform approach to searching protoclusters on which this study builds enables us to link observed variations at the onset of cosmic noon to different pathways in the history of cluster formation.
\end{abstract}


\section{Introduction}
Galaxy clusters are the most massive objects in the Universe, pinpointing where mass density has drastically risen
from the small fluctuations of the initial condition of the Universe \citep[e.g.,][]{peebles80}.
Such extremely high-density environments are rare and tend to be formed at the nodes of the cosmic web
\citep[e.g.,][]{libeskind18,malavasi20,cohn22}; thus, galaxy clusters are key components of the large-scale
structure of the Universe.
In addition to their use as cosmological probes, galaxy clusters play a key role in shaping galaxy evolution
as galaxies residing in high-density environments exhibit distinct properties from field galaxies in the local
Universe \citep[e.g.,][]{dressler80,balogh99,gallazzi21}.
Galaxies are far from ``closed-box'' systems and involve complicated physical interactions with their surrounding
environments \citep[e.g.][]{somerville15,alberts22}.
Along the filaments of the cosmic web, mass and gas accrete onto a galaxy and will fuel further star formation.
Galaxies can gain stellar mass by merging with another one instead of forming stars in-situ.
On the other hand, galaxy mergers/interactions can also lead to quenching of star-forming activity, especially by gas stripping
from low-mass galaxies or by promoting gas consumption and outflows.
It is also possible that these physical processes can occur repeatedly or simultaneously.
Although the consequences of such accumulated environmental effects on galaxy evolution are imprinted onto mature galaxy
clusters as seen in the local Universe, complementary insights on these processes can be gained by observing their developmental phases at high
redshifts, so-called ``protoclusters''. Their study helps to understand how and when cluster galaxies differentiated from
field counterparts.

Due to the rarity of protoclusters, various methods are used to boost the chance of discoveries.
One of the most effective ways is targeting a signpost of overdensity.
Radio galaxies (RGs) and quasars (QSOs) are expected to be formed in dense environments, and their high luminosity
powered by active galactic nuclei (AGNs) make them easy to identify. 
Based on this technique, many protoclusters at $z>2$ are successfully discovered \citep[e.g.,][]{venemans07,wylezalek13,luo22}.
In a similar manner, protoclusters are found around bright sub-mm galaxies (SMGs)
\citep[e.g.,][]{miller18,oteo18,calvi23}.
However, such galaxies are also found in more average- or lower-density regions
\citep[e.g.,][]{uchiyama18,noirot18,cheng20}.
\citet{hatch14} confirmed that RGs are preferentially located in high-density environments, but there is no clear correlation between overdensity and the physical properties of RGs.
Furthermore, \citet{suzuki24} suggested that strong UV radiation from a QSO suppresses the star-forming activity of
surrounding low-mass galaxies based on the deficit of number density of faint galaxies in the vicinity of QSOs.
\citet{yang18} suggest that black-hole growth is mainly controlled by the stellar mass of host galaxies, regardless of environments where they are in.
The many successes of protocluster discoveries targeting AGNs may be created by an indirect correlation through massive galaxies.
Therefore, the connection between protoclusters and the presence of such peculiar galaxies is still under debate,
and the sample of protoclusters selected by using signposts may be just a part of the whole population.

In addition to the various methods of protocluster search, protoclusters exhibit the large diversity of their properties.
Some protoclusters include many starburst galaxies, and their total SFR exceeds $10^4\,\mathrm{M_{\sun}\,yr^{-1}}$ \citep[e.g.,][]{hill20,wang25}.
Others are dominated by quiescent galaxies \citep[e.g.,][]{mcconachie22,tanaka24}. 
AGN fraction in protoclusters also ranges from lower to higher than that in field \citep[e.g.,][]{macuga19,shah25}.
Furthermore, we can see contradictory results in other features, such as metallicity \citep[e.g.,][]{calabro22,yang26,wang26} and internal structure \citep[e.g.,][]{toshikawa20,shi21}.
As protoclusters at $z\sim2\mathrm{-}3$ are expected to be at the transition phase of cluster formation \citep{daddi22}, such diversity could, in principle, provide key hints about their developmental stages.
However, the heterogeneous sample of protoclusters due to observational biases makes it difficult to fairly compare each of them. 
So far, confirmed protoclusters tend to be biased toward rich structures.
The SSA22 and Spiderweb protoclusters, which are the most investigated samples at $z\sim2\mathrm{-}3$, are expected to grow into $>10^{15}\,\mathrm{M_{\sun}}$ halo by $z=0$ \citep{topping18,mascolo23}.
Another unique example of SPT2349-56 protocluster, which has a compact core composed of dusty starburst galaxies, is also expected to be the progenitor of $>10^{15}\,\mathrm{M_{\sun}}$ halo at $z=0$.
As shown in Figure~7 of \citet{overzier16}, the distribution of expected descendant halo mass at $z=0$ of high-redshift protoclusters is skewed to the massive end, and some protoclusters would be more massive than any clusters in the local universe.
Although the estimate of descendant halo mass involves large uncertainty, selection bias would be one of the reasons of such a skewed distribution.

To overcome this situation, blind surveys are required so as to identify the rare protocluster objects without
signposts as a prior.
Although large observing time had to be devoted, a few protoclusters were serendipitously discovered, showing
that protoclusters do not always include RGs or QSOs \citep[e.g.,][]{ouchi05,douglas10,toshikawa12}. 
As such searches for protoclusters were expanded into wider or deeper datasets, the number of protoclusters has steadily been
increased \citep[e.g.,][]{chiang14,ramakrishnan24,sutanto26}.
It should be noted that even blind surveys rely on a certain galaxy population to measure overdensity, which will introduce a source of potential observational bias.
There are a few examples of protoclusters which are investigated by multiple galaxy populations, such as young
star-forming galaxies, massive galaxies, or dusty galaxies.
\citet{rotermund21} reported that a protocluster at $z=4.3$ is dominated by dusty galaxies and sparsely populated
by Lyman break galaxies (LBGs).
Even if a protocluster exhibits overdensities with multiple galaxies populations, the spatial distributions kf
each galaxy population within a protocluster can be different, or the properties of protocluster galaxies depend
on radial distance from the protocluster centre \citep[e.g.,][]{koyama13,shi25}.
These unique features reflect the physical processes of galaxy evolution in protoclusters, but observational
biases or inhomogeneity frequently hamper detailed comparisons among protocluster samples.

In recent years, protocluster searches have been significantly expanded by the advents of new instruments/facilities.
The James Webb Space Telescope (JWST) confirmed protoclusters at $z\gtrsim6$ \citep[e.g.,][]{hashimoto23,morishita23,martin26}, and a candidate was found even at $z\sim10$ \citep[e.g.,][]{scholtz24,wu26}.
In spite of small FoV, JWST's deep near-infrared (NIR) imaging makes it possible to map out the large-scale structure of the Universe across a wider redshift range based on the accurate estimate of photometric redshifts \citep[e.g.,][]{hatamnia25,sutanto26}.
In addition to such deep NIR observations, wide-field optical imaging from ground-based telescopes can carry out a systematic search for protoclusters at the same epoch, allowing us to notice the large diversity of protoclusters.
The Subaru Strategic Programme with Hyper SuprimeCam \citep[HSC-SSP:][]{aihara18} is, so far, the largest survey performed by an 8m-class telescope.
The HSC enables to perform optical imaging with $1.5\,\mathrm{deg}$-diameter FoV.
By using this wide-field imaging capability, the HSC-SSP conducted an unprecedented wide and deep survey, which is composed of three layers: Wide ($\sim1,000\,\mathrm{deg^2}$, $i$-band depth of $m_i\sim26.0\,\mathrm{mag}$), Deep ($\sim25\,\mathrm{deg^2}$, $m_i\sim26.5\,\mathrm{mag}$), and Ultradeep ($\sim3\,\mathrm{deg^2}$, $m_i\sim27.0\,\mathrm{mag}$).
In the Wide layer, \citet{toshikawa18} identified $N\sim200$ protocluster candidates at $z\sim4$, enabling statistical studies of protocluster properties \citep[e.g.,][]{uchiyama18,kubo19,ito19}.
Our sample of protocluster candidates is expected to be the progenitors of $\sim4\times10^{14}\,\mathrm{M_{\sun}}$ halo at $z=0$ by clustering analysis; thus, the HSC-SSP was able to select the representative sample of typical protoclusters compared with previous studies targeting a few extreme protoclusters.
A sample of protocluster candidates across a wider redshift range was constructed in the Deep/Ultradeep (DUD) layer \citep{toshikawa24},
which is composed of $N>100$ protocluster candidates at $z\sim3\mathrm{-}5$.
Thanks to the consistent selection technique on top of the uniform dataset, we will be able to look into the physical nature of protoclusters with little impact of observational biases.
We have found an overabundance of rest-UV bright galaxies in high-density regions across $z\sim3\mathrm{-}5$.
In the DUD layer, the deep near-infrared data of $J$-, $H$-, and $K(K_S)$-bands are also available.
Galaxies in high-density environments would have comparable star-forming activity, or specific SFR ($sSFR=SFR/M_*$), to coeval field galaxies according to SED fitting.
Thus, while protocluster galaxies are expected to follow the consistent star-formation history with field galaxies, they would be formed earlier.
This earlier galaxy formation in high-density environments causes the increment of stellar mass (UV brightness) at an observed epoch; in other words, the cosmic clock is ahead in protoclusters.
This systematic sample enables us to notice a difference between protocluster and field galaxies at high redshifts, but photometric candidates are heavily contaminated by foreground/background galaxies.
Even within a single protocluster, galaxy properties can vary and possibly show radial dependency as predicted by the ``Inside-Out Growth'' model \citep{chiang17}.

Our pilot follow-up spectroscopy confirmed the existence of a protocluster at $z=3.70$ \citep{toshikawa25}.
Although the average of physical properties (e.g., UV or Ly$\alpha$ luminosity) of its member galaxies is comparable to that of field galaxies, the precise measurement of three-dimensional distribution by spectroscopy found that the physical properties of member galaxies are correlated with the separation from nearest neighbours rather than the radial distance from the protocluster centre, implying that it is still in an early phase of cluster formation.
In contrast, it is also true that mature protoclusters are found at $z\gtrsim3$.
Even at the same cosmic epoch, protoclusters would be on various developmental phases possibly depending on their descendant halo mass at $z=0$.
It is also natural that galaxy clusters following different pathways will be eventually a similar scale of structures.
However, these inherent diversity can be enshrouded by inhomogeneous samples selected by various techniques.
Therefore, we have expanded our follow-up spectroscopic campaign on our systematic sample of protocluster candidates.
This study newly observes three protocluster candidates at $z\sim3$ and compared them not only with coeval field galaxies but also among the protoclusters.
Since the targets of this study are selected from the same dataset using identical techniques, the diversity of protoclusters, if any, can be attributed to the physical nature of protoclusters, allowing us to investigate the link between cluster formation and galaxy evolution

The structure of this paper is as follows.
Section~\ref{sec:obs} describes the details of target protocluster candidates and follow-up spectroscopic
observation.
The confirmation of protoclusters and their physical properties are presented in Section~\ref{sec:res}.
We discuss the developmental phase of protoclusters based on the stellar population and Ly$\alpha$ nature of member
galaxies in Section~\ref{sec:dis}.
Finally, our conclusions and their implications are summarized in Section~\ref{sec:con}.
The following cosmological parameters are assumed: $\Omega_M=0.3$, $\Omega_\Lambda=0.7$, and
$H_0=70\,\mathrm{km\,s^{-1}\,Mpc^{-1}}$.
Magnitudes are given in the AB system.

\section{Observations} \label{sec:obs}
\subsection{Targets}
The systematic sample of protocluster candidates at $z\sim3\mathrm{-}5$ was constructed based on the wide-field
surveys of HSC-SSP \citep{aihara18} and CLAUDS \citep{sawicki19}.
These surveys provide deep multi-band images from $U$- to $y$-bands over an area of $\sim20\,\mathrm{deg^2}$,
enabling to select dropout galaxies down to $M^*_\mathrm{UV}+1.0$ ($M^*_\mathrm{UV}$) at $z\sim3$ ($z\sim5$) where
$M^*_\mathrm{UV}$ is the characteristic magnitude of the luminosity function \citep[cf.,][]{schechter76,bouwens21,harikane22}.
Dropout galaxies from $z\sim3$ ($U$-dropout) to $z\sim5$ ($r$-dropout) are selected by the same colour selections as those of \citet{ono18}.
Then, the local number density of dropout galaxies is determined by counting galaxies within a fixed aperture of
$0.75\,\mathrm{proper\>Mpc}$ (pMpc) radius for all redshift samples.
The regions exhibiting $>4\sigma$ overdensity at their peak are regarded as protocluster candidates.
It should be noted that the redshift window of the dropout technique ($\Delta z\sim1$) is much larger than the size of
protocluster ($\Delta z\sim0.05$).
The signal of surface overdensity from a protocluster can thus be altered by projection effects.
According to the model prediction of \citet{toshikawa16}, $\sim90\%$ of candidates are expected to include genuine
protoclusters, while some of them ($\sim5\mathrm{-}15\%$) may not be feasible to observationally confirm due to the weakness of their 
overdensity signal.
In total, 111 protocluster candidates are selected from $z\sim3$ to $z\sim5$.
We refer the reader to \citet{toshikawa24} for details on the identification of protocluster candidates in the DUD
layer of the HSC-SSP.

\begin{figure*}
\includegraphics[width=\hsize]{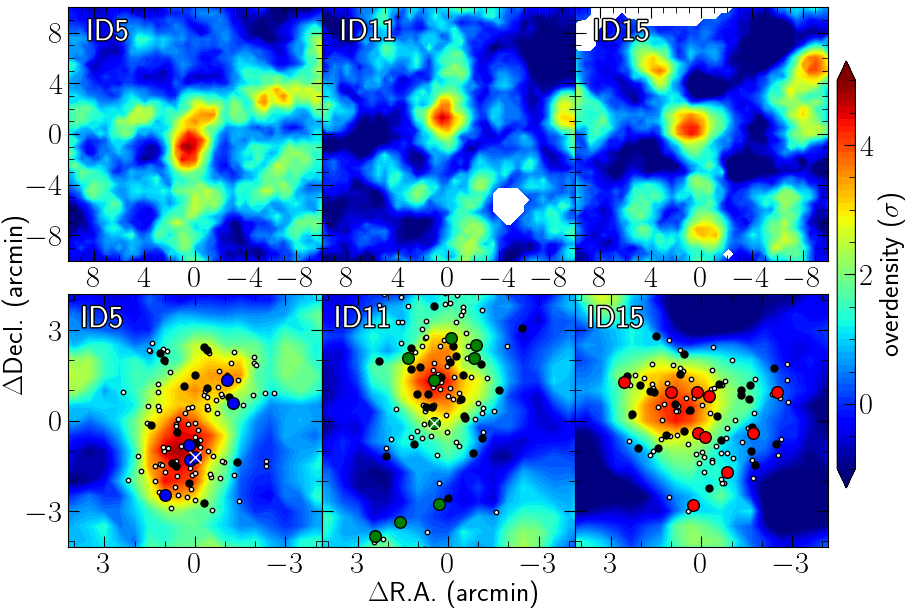}
\caption{Overdensity maps of $U$-dropout galaxies in the three target regions of protocluster candidates. The top and bottom panels show wider area and close-up of targets, respectively. The white open points represent spectroscopically-observed $U$-dropout galaxies that do not exhibit strong emission lines. $U$-dropout galaxies with Ly$\alpha$ emission are shown by the filled points. In the protocluster candidates of ID5, ID11 and ID15, confirmed protocluster members are indicated by the blue, green and red filled points, respectively, while field galaxies are the black filled points. The white crosses show AGNs in protoclusters. Noted that in total four AGNs are identified in this study, but two AGNs in field are not plotted on the figure. The origins of left (ID5), middle (ID11), and right (ID15) panels correspond to (R.A., Decl.) $=$ (09:58:55.1, $+$01:33:32.8), (23:27:27.2, $-$00:43:27.8), and (16:09:08.5, $+$55:32:37.5), respectively.}
\label{fig:cntr}
\end{figure*}
Among this sample of protocluster candidates, this study focuses on $U$-dropout ($z\sim3.1$) ones because such
lower-$z$ sample makes it more feasible to constrain the physical properties of each protocluster's member galaxies.
The DUD layer is composed of four separate fields: COSMOS ($\mathrm{RA}\sim150\,\mathrm{deg}$), DEEP2-3
($\mathrm{RA}\sim352\,\mathrm{deg}$), ELAIS-N1 ($\mathrm{RA}\sim243\,\mathrm{deg}$), and XMM-LSS
($\mathrm{RA}\sim36\,\mathrm{deg}$).
We have put a higher priority on protocluster candidates with higher overdensity for our follow-up observation.
In addition, protocluster candidates located near the edge of the fields are not included among the list of possible targets
regardless of their overdensity, since protocluster galaxies can be distributed widely over angular scales of
$\sim15\mathrm{-}30\,\mathrm{arcmin}$ \citep{muldrew15,chiang17}.
Finally, due to the target visibility on the allocated nights, we have taken spectroscopy on the three protocluster candidates of $U$-dropout galaxies: ID5 in COSMOS, ID11 in DEEP2-3, and ID15 in ELAIS-N1 (the IDs of protocluster candidates are the same as in Table~2 of \citealt{toshikawa24}).
Figure~\ref{fig:cntr} shows the overdensity contours of $U$-dropout galaxies in the target protocluster candidates.
Although the outskirts, or relatively lower-density regions, could be susceptible to projection effects, each target has a unique morphology, like elongated or circular.
The number of $U$-dropout galaxies enclosed in a $0.75\,(1.5)\,\mathrm{pMpc}$-radius aperture is 131 (455), 129 (419), and 153 (479) down to the $5\sigma$ limiting magnitude of $i$-band in the protocluster candidates of ID5, ID11, and ID15, respectively.
Our target protocluster candidates having $\sim4.3\sigma$-$4.8\sigma$ overdensity significance are in about twice denser regions than the average.
From a larger viewpoint, some of them seem to be accompanied by lower ($\sim2\mathrm{-}3\sigma$) overdense regions, and the others would be isolated.
This variation may reflect the different developmental stages or pathways of cluster formation.

\begin{table*}
\caption{Overview of the follow-up spectroscopy}
\label{tab:obs}
\begin{tabular}{ccccccc}
\hline
PCL & Date & Mask & ${N_\mathrm{slit}}^a$ & Grism & ${t_\mathrm{exp}}$ & ${\sigma_\mathrm{line}}^b$ \\
 & & & & & (min) & ($\mathrm{erg\,s^{-1}\,cm^{-2}}$) \\
\hline
ID5 & 15th April 2023 & A & 30 & 300B & 60 & $3.9\times10^{-19}$ \\
 & & B & 28 & 300B & 60 & $2.3\times10^{-18}$ \\
 & & C & 28 & 300B & 60 & $1.4\times10^{-18}$ \\
 & 5th March 2024 & D & 28 & VPH520 & 140 & $1.6\times10^{-18}$ \\
ID11 & 15th July 2023 & A & 30 & VPH520 & 120 & $2.4\times10^{-19}$ \\
 & & B & 27 & VPH520 & 80 & $4.3\times10^{-19}$ \\
 & 16th July 2023 & C & 29 & VPH520 & 180 & $8.2\times10^{-19}$ \\
 & 4th September 2023 & D & 14 & VPH520 & 180 & $1.6\times10^{-18}$ \\
 & 4th September 2023 & E & 12 & VPH520 & 120 & $2.1\times10^{-18}$ \\
 & 5,6th September 2023 & F & 8 & VPH520 & 300 & $1.7\times10^{-18}$ \\
 & 6th September 2023 & G & 11 & VPH520 & 160 & $2.2\times10^{-18}$ \\
ID15 & 15th April 2023 & A & 28 & 300B & 60 & $4.6\times10^{-18}$ \\
 & & B & 29 & 300B & 60 & $4.4\times10^{-19}$ \\
 & & C & 29 & 300B & 80 & $5.5\times10^{-19}$ \\
 & 15th July 2023 & A & 27 & VPH520 & 60 & $7.1\times10^{-19}$ \\
 & & B & 28 & VPH520 & 60 & $8.4\times10^{-19}$ \\
 & & C & 28 & VPH520 & 60 & $1.4\times10^{-18}$ \\
 & 5th March 2024 & D & 29 & VPH520 & 180 & $7.4\times10^{-19}$ \\
\hline
\multicolumn{7}{l}{$^a$ The number of slits which can be used for analysis. Since the VPH grism produces a shifted image, a few slits may fall outside the detector.} \\
\multicolumn{7}{l}{$^b$ The $1\sigma$ detection limit of emission lines. Its FWHM is assumed to be $\sim5\,${\AA}.} \\
\end{tabular}
\end{table*}
\subsection{Follow-up spectroscopy}
The accurate determination of redshifts by spectroscopy allows us to evaluate clustering strength along the
redshift dimension, which will serve as direct evidence of the existence of a protocluster.
Our follow-up spectroscopy was performed by Subaru/FOCAS \citep{kashikawa02} in 2023 and 2024 (proposal ID: S23A089,
S23B007, and S24A098).
The high-density nature of protoclusters allows optimal use of its multi-object spectroscopy (MOS) capabilities.
The FoV of FOCAS is $6\,\mathrm{arcmin}$ diameter, and nearly 30 slits can be allocated within the FoV.
We use the 300B or VPH520 grisms together with a $0.8\,\mathrm{arcsec}$ slit width, resulting in a spectral resolution of $R\sim500$ or 1500, respectively.
The typical wavelength coverages of 300B and VPH520 grisms are $4300\mathrm{-}8000\,${\AA} and $4400\mathrm{-}6000\,${\AA}, respectively.
The 300B grism tends to have a higher sensitivity especially to broad emission lines, while the higher spectral
resolution of the VPH520 grism can subtract sky emissions more accurately and detect faint narrow emission lines even in the vicinity of strong sky lines.
Due to the large number of $U$-dropout galaxies in the target regions, each protocluster candidate is observed by
several MOS masks.
Our follow-up spectroscopy was conducted over multiple runs, and the Subaru telescope was operated by in classical
mode (i.e., with observing nights being fixed irrespective of sky conditions).
As the sky conditions of some allocated nights were not good, we made the total on-source exposure times flexible
in order to achieve the required flux limit.
The details of our follow-up spectroscopy are summarised in Table~\ref{tab:obs}.
In total, $N=112$, 117, 112 $U$-dropout galaxies are spectroscopically observed in the protocluster candidates of
ID5, ID11, and ID15, respectively.
The fraction of slit allocated $U$-dropout galaxies among all in the FoVs of our follow-up spectroscopy is about 30\%.
Although there is no tight correlation between the Ly$\alpha$ flux and UV continuum of dropout
galaxies, brighter galaxies in $i$-band received higher priority for slit allocation.
The combination of small slit-allocation fraction and weight on brighter galaxies causes a significant difference of luminosity distribution between the photometric sample of dropout galaxies and spectroscopic targets.
Due to the large redshift window of $U$-dropout galaxies, our follow-up spectroscopy does not identify only protocluster members but also foreground/background galaxies, which can be the field counterparts of protocluster members to discuss possible environmental differences with less observational biases.

We reduced the FOCAS data using standard procedures with IRAF, including bias subtraction, flat fielding, wavelength
calibration, sky subtraction, combining two-dimensional spectra, and flux calibration.
The Galactic extinction is corrected by the extinction map of \citet{schlafly11}.
It should be noted that the Galactic extinction over the three regions is almost consistent (the difference between maximum and minimum is $\Delta A_V\sim0.1\,\mathrm{mag}$).
The wavelength solution derived from an arc lamp (ThAr) frame was applied to the science frames.
Such calibration data were taken at the beginning or end of an observing night, so slight wavelength shifts between
science and comparison frames were corrected by using a bright sky line of [O{\sc i}] at $\lambda=5577.34\,${\AA}.
It should be noted that strong sky lines (e.g., N, O, and Na) are too sparse over the whole wavelength coverage of our spectroscopy to determine the relation between pixel and wavelength.
Each MOS mask includes a bright star ($m_g\sim21\mathrm{-}22\,\mathrm{mag}$), which can be utilised for monitoring
the variation of sky conditions (e.g., seeing size or transparency) between exposures.
Finally, we searched for possible emission lines by following the same approach as outlined in Sec.~2.2 of \citet{toshikawa25}.
All single emission lines are regarded as Ly$\alpha$ emissions, since the possible contaminants from lower-redshift
galaxies (e.g., H$\beta$, [O{\sc ii}], or [O{\sc iii}] emissions) will be detected as multiple lines.
In the case of the VPH520 grism, its wavelength coverage is not sufficiently wide to detect all multiple lines of
[O{\sc ii}] and the others; however, its resolving power is sufficiently high to differentiate a [O{\sc ii}] doublet from
a Ly$\alpha$ singlet.
Most of emission lines show redward skewed line profiles, which is a unique feature of Ly$\alpha$ emission from
high-redshift galaxies.
Finally, we have spectroscopically determined the redshifts of $N=22$, 34, and 37 $U$-dropout galaxies in the
protocluster candidates of ID5, ID11, and ID15, respectively.
As it is not always that dropout (continuum-selected) galaxies have strong Ly$\alpha$ emissions to be detected by our spectroscopy, the redshifts of some targets cannot be determined even if they are genuine high-redsfhit galaxies.
It should be noted that this may cause an observation bias on the study of galaxy population in protoclusters.
Table.~\ref{tab:spec} describes the observed properties of each of the spectroscopically-confirmed galaxies.
Redshifts are determined by wavelength at the peak of Ly$\alpha$ emission, and Ly$\alpha$ flux,
$f_\mathrm{Ly\alpha}$, is quantified by integrating flux within a line profile.
In this estimate, the absorption of intergalactic medium (IGM) is not corrected.
Since our follow-up spectroscopy is too shallow to detect the UV continuum of target $U$-dropout galaxies, their $r$-band magnitude is used for the calculation of absolute UV magnitude, $M_\mathrm{UV}$, as well as rest-frame equivalent width, $EW_0$, by assuming a flat UV slope ($f_\lambda\propto\lambda^\beta$, $\beta=-2$).
\begin{figure*}
\includegraphics[width=\hsize]{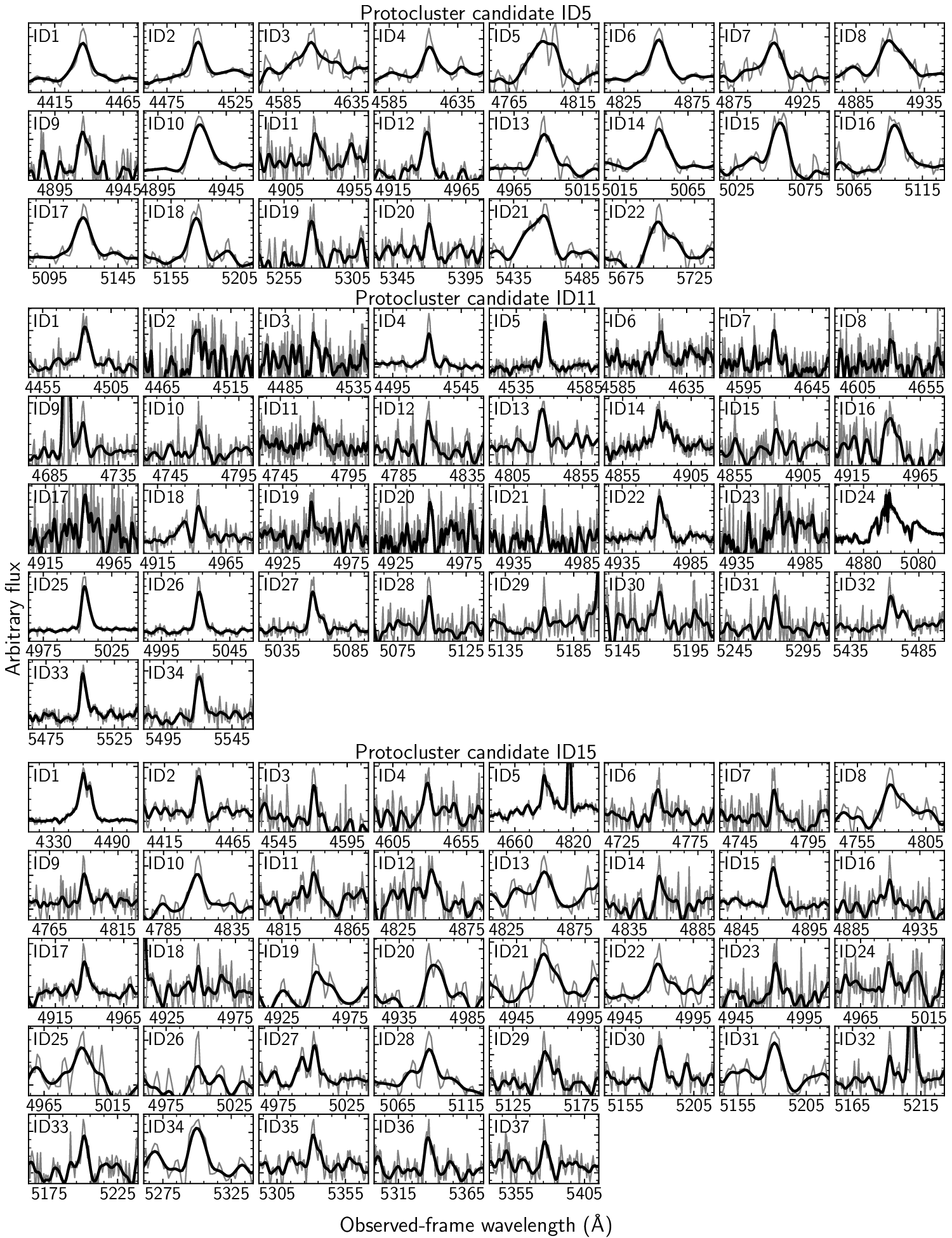}
\caption{One-dimensional spectra of all $U$-dropout galaxies confirmed by our spectroscopy. The black and gray lines are spectra with and without smoothing by Gaussian kernel ($\mathrm{FWHM=5}\,\mathrm{pixel}$).}
\label{fig:sp1d}
\end{figure*}

\begin{table*}
\caption{Properties of spectroscopically-confirmed $U$-dropout galaxies.}
\label{tab:spec}
\begin{tabular}{cccccccccc}
\hline
No. & R.A. & Decl. & $m_r$ & $z$ & $M_\mathrm{UV}$ & $f_\mathrm{Ly\alpha}$ & $L_\mathrm{Ly\alpha}$ & $EW_0$ & $S_w^a$ \\
 & (J2000) & (J2000) & (mag) & & (mag) & ($10^{-18}\,\mathrm{erg\,s^{-1}\,cm^{-2}}$) & ($10^{42}\,\mathrm{erg\,s^{-1}}$) & (\AA) & (\AA) \\
\hline
\multicolumn{10}{c}{Protocluster candidate of ID5} \\
1 & 09:58:53.94 & +01:35:59.8 & $25.11\pm0.03$ & $2.649$ & $-19.95\pm0.03$ & $32.77\pm1.87$ & $1.88\pm0.11$ & $22.3\pm1.5$ & $6.7\pm2.4$ \\
2 & 09:58:57.50 & +01:33:09.7 & $23.93\pm0.02$ & $2.700$ & $-21.35\pm0.01$ & $19.30\pm1.13$ & $1.16\pm0.07$ & $3.8\pm0.2$ & $5.5\pm2.5$ \\
3 & 09:58:55.84 & +01:31:29.3 & $24.62\pm0.05$ & $2.790$ & $-19.90\pm0.05$ & $26.86\pm7.20$ & $1.74\pm0.47$ & $21.7\pm5.9$ & $2.8\pm13.0$ \\
4 & 09:58:56.54 & +01:34:41.1 & $23.94\pm0.02$ & $2.795$ & $-21.29\pm0.01$ & $14.99\pm1.24$ & $0.98\pm0.08$ & $3.4\pm0.3$ & $8.4\pm2.3$ \\
5 & 09:59:00.97 & +01:33:24.6 & $24.95\pm0.05$ & $2.940$ & $-20.52\pm0.03$ & $23.53\pm3.90$ & $1.73\pm0.29$ & $12.2\pm2.0$ & $4.8\pm21.6$ \\
6 & 09:58:55.13 & +01:35:03.4 & $24.52\pm0.04$ & $2.990$ & $-20.85\pm0.02$ & $22.33\pm1.24$ & $1.71\pm0.10$ & $8.9\pm0.5$ & $4.8\pm3.6$ \\
7 & 09:58:55.89 & +01:32:43.7 & $25.41\pm0.08$ & $3.034$ & $-20.27\pm0.04$ & $32.02\pm5.43$ & $2.54\pm0.43$ & $22.6\pm3.9$ & $16.4\pm10.0$ \\
8 & 09:58:59.03 & +01:31:05.7 & $24.09\pm0.02$ & $3.039$ & $-20.82\pm0.03$ & $62.27\pm7.12$ & $4.96\pm0.57$ & $26.4\pm3.1$ & $14.9\pm7.2$ \\
9 & 09:58:55.74 & +01:32:15.2 & $25.40\pm0.07$ & $3.044$ & $-19.63\pm0.14$ & $14.08\pm7.09$ & $1.13\pm0.57$ & $17.9\pm9.3$ & $1.6\pm3.5$ \\
10 & 09:58:55.06 & +01:32:19.5 & $23.41\pm0.01$ & $3.051$ & $-21.97\pm0.01$ & $58.10\pm1.93$ & $4.68\pm0.16$ & $8.7\pm0.3$ & $5.9\pm1.7$ \\
11 & 09:58:50.83 & +01:34:53.5 & $25.43\pm0.09$ & $3.051$ & $-20.07\pm0.09$ & $14.50\pm7.95$ & $1.17\pm0.64$ & $12.4\pm6.9$ & $6.8\pm3.9$ \\
12 & 09:58:50.06 & +01:34:06.9 & $25.13\pm0.05$ & $3.064$ & $-20.25\pm0.08$ & $29.20\pm8.35$ & $2.38\pm0.68$ & $21.4\pm6.4$ & $0.7\pm6.8$ \\
13 & 09:58:53.55 & +01:34:38.0 & $25.42\pm0.11$ & $3.102$ & $-19.70\pm0.07$ & $4.97\pm0.86$ & $0.42\pm0.07$ & $6.2\pm1.1$ & $0.7\pm10.4$ \\
14 & 09:58:49.45 & +01:32:09.5 & $25.76\pm0.09$ & $3.149$ & $-19.36\pm0.09$ & $19.24\pm1.57$ & $1.67\pm0.14$ & $34.1\pm4.2$ & $2.7\pm2.6$ \\
15 & 09:58:53.89 & +01:30:48.1 & $25.26\pm0.07$ & $3.156$ & $-20.38\pm0.04$ & $9.37\pm1.64$ & $0.82\pm0.14$ & $6.6\pm1.2$ & $-3.9\pm7.1$ \\
16 & 09:58:50.58 & +01:34:20.5 & $24.93\pm0.05$ & $3.188$ & $-20.50\pm0.04$ & $33.72\pm3.82$ & $3.02\pm0.34$ & $21.5\pm2.5$ & $21.2\pm6.3$ \\
17 & 09:58:57.55 & +01:32:02.9 & $25.53\pm0.07$ & $3.211$ & $-20.19\pm0.05$ & $40.92\pm3.51$ & $3.72\pm0.32$ & $35.6\pm3.5$ & $5.6\pm6.0$ \\
18 & 09:58:59.17 & +01:35:32.4 & $24.74\pm0.03$ & $3.257$ & $-20.46\pm0.04$ & $3.14\pm0.83$ & $0.30\pm0.08$ & $2.2\pm0.6$ & $1.6\pm30.5$ \\
19 & 09:58:58.08 & +01:32:09.2 & $25.84\pm0.13$ & $3.341$ & $-19.82\pm0.16$ & $7.59\pm4.45$ & $0.76\pm0.45$ & $10.2\pm6.2$ & $1.2\pm4.6$ \\
20 & 09:58:54.96 & +01:32:47.8 & $25.88\pm0.13$ & $3.416$ & $-20.34\pm0.11$ & $8.91\pm4.82$ & $0.94\pm0.51$ & $7.8\pm4.3$ & $2.4\pm4.5$ \\
21 & 09:58:59.75 & +01:35:47.7 & $26.20\pm0.09$ & $3.489$ & $-19.49\pm0.13$ & $9.42\pm1.59$ & $1.05\pm0.18$ & $19.0\pm4.0$ & $-9.7\pm8.6$ \\
22 & 09:58:53.51 & +01:35:53.7 & $25.88\pm0.10$ & $3.688$ & $-19.94\pm0.12$ & $4.55\pm0.86$ & $0.58\pm0.11$ & $6.9\pm1.5$ & $8.8\pm8.3$ \\
\hline
\multicolumn{10}{c}{Protocluster candidate of ID11} \\
1 & 23:27:26.49 & -00:40:59.1 & $25.30\pm0.06$ & $2.689$ & $-20.02\pm0.07$ & $22.00\pm3.33$ & $1.31\pm0.20$ & $14.5\pm2.4$ & $3.1\pm8.6$ \\
2 & 23:27:28.09 & -00:42:00.7 & $25.32\pm0.06$ & $2.695$ & $-19.72\pm0.09$ & $17.99\pm5.63$ & $1.07\pm0.34$ & $15.8\pm5.2$ & $-0.8\pm1.4$ \\
3 & 23:27:25.48 & -00:42:44.7 & $24.79\pm0.05$ & $2.706$ & $-20.16\pm0.06$ & $10.31\pm4.36$ & $0.62\pm0.26$ & $6.1\pm2.6$ & $3.9\pm2.3$ \\
4 & 23:27:25.25 & -00:41:56.8 & $25.09\pm0.07$ & $2.719$ & $-20.08\pm0.07$ & $23.29\pm1.56$ & $1.42\pm0.10$ & $15.0\pm1.4$ & $0.8\pm0.9$ \\
5 & 23:27:30.97 & -00:41:41.5 & $25.74\pm0.11$ & $2.747$ & $-19.75\pm0.09$ & $54.50\pm7.84$ & $3.41\pm0.49$ & $48.7\pm8.3$ & $7.3\pm1.8$ \\
6 & 23:27:29.39 & -00:43:00.6 & $26.24\pm0.13$ & $2.796$ & $-19.03\pm0.18$ & $23.60\pm6.52$ & $1.54\pm0.43$ & $42.7\pm14.1$ & $4.9\pm1.3$ \\
7 & 23:27:17.40 & -00:40:24.1 & $23.91\pm0.02$ & $2.798$ & $-21.09\pm0.03$ & $6.41\pm2.60$ & $0.42\pm0.17$ & $1.7\pm0.7$ & $3.2\pm2.1$ \\
8 & 23:27:29.82 & -00:42:34.4 & $25.35\pm0.07$ & $2.807$ & $-19.90\pm0.09$ & $8.50\pm4.37$ & $0.56\pm0.29$ & $7.0\pm3.6$ & $0.8\pm1.4$ \\
9 & 23:27:35.43 & -00:44:37.9 & $26.77\pm0.20$ & $2.874$ & $-19.00\pm0.20$ & $7.15\pm1.95$ & $0.50\pm0.14$ & $14.2\pm4.8$ & $8.0\pm59.0$ \\
10 & 23:27:31.71 & -00:44:14.3 & $25.21\pm0.06$ & $2.920$ & $-20.02\pm0.08$ & $3.80\pm1.22$ & $0.28\pm0.09$ & $3.1\pm1.0$ & $0.1\pm2.5$ \\
11 & 23:27:24.97 & -00:43:22.5 & $24.15\pm0.02$ & $2.923$ & $-21.03\pm0.03$ & $12.63\pm4.05$ & $0.92\pm0.29$ & $4.0\pm1.3$ & $4.5\pm2.5$ \\
12 & 23:27:32.12 & -00:41:45.1 & $24.55\pm0.04$ & $2.954$ & $-20.67\pm0.05$ & $4.05\pm1.12$ & $0.30\pm0.08$ & $1.8\pm0.5$ & $3.4\pm1.9$ \\
13 & 23:27:29.06 & -00:39:38.9 & $24.26\pm0.02$ & $2.971$ & $-21.02\pm0.04$ & $10.49\pm1.03$ & $0.79\pm0.08$ & $3.5\pm0.4$ & $7.3\pm1.9$ \\
14 & 23:27:27.20 & -00:46:01.2 & $23.51\pm0.01$ & $3.016$ & $-21.51\pm0.03$ & $57.68\pm7.02$ & $4.51\pm0.55$ & $12.8\pm1.6$ & $12.7\pm1.9$ \\
15 & 23:27:32.18 & -00:41:03.4 & $24.93\pm0.04$ & $3.018$ & $-20.47\pm0.06$ & $3.26\pm0.98$ & $0.26\pm0.08$ & $1.9\pm0.6$ & $2.6\pm1.5$ \\
16 & 23:27:32.53 & -00:41:23.6 & $24.55\pm0.03$ & $3.066$ & $-20.79\pm0.05$ & $6.26\pm1.53$ & $0.51\pm0.12$ & $2.8\pm0.7$ & $6.4\pm7.1$ \\
17 & 23:27:28.36 & -00:46:14.5 & $24.21\pm0.03$ & $3.066$ & $-20.99\pm0.04$ & $10.76\pm5.37$ & $0.88\pm0.44$ & $4.0\pm2.0$ & $6.2\pm3.8$ \\
18 & 23:27:23.46 & -00:40:57.1 & $25.18\pm0.07$ & $3.068$ & $-20.40\pm0.07$ & $12.47\pm1.54$ & $1.02\pm0.13$ & $8.0\pm1.1$ & $6.6\pm3.8$ \\
19 & 23:27:23.72 & -00:41:23.3 & $24.18\pm0.03$ & $3.070$ & $-20.66\pm0.06$ & $4.92\pm1.72$ & $0.40\pm0.14$ & $2.5\pm0.9$ & $3.4\pm1.3$ \\
20 & 23:27:26.78 & -00:40:43.0 & $25.95\pm0.11$ & $3.071$ & $-19.27\pm0.19$ & $4.39\pm2.96$ & $0.36\pm0.24$ & $8.0\pm5.6$ & $1.9\pm2.3$ \\
21 & 23:27:29.11 & -00:42:07.1 & $25.96\pm0.11$ & $3.078$ & $-19.54\pm0.15$ & $4.04\pm2.35$ & $0.33\pm0.19$ & $5.8\pm3.5$ & $3.3\pm2.6$ \\
22 & 23:27:33.55 & -00:46:50.4 & $24.31\pm0.02$ & $3.082$ & $-21.06\pm0.04$ & $55.25\pm6.94$ & $4.56\pm0.57$ & $19.5\pm2.6$ & $12.0\pm1.9$ \\
23 & 23:27:36.87 & -00:47:17.8 & $24.64\pm0.04$ & $3.082$ & $-20.46\pm0.07$ & $14.43\pm7.11$ & $1.19\pm0.59$ & $8.9\pm4.4$ & $5.9\pm5.0$ \\
24 & 23:27:29.05 & -00:43:33.3 & $22.10\pm0.00$ & $3.093$ & $-23.05\pm0.01$ & $1842.00\pm16.00$ & $153.19\pm1.33$ & $104.4\pm1.2$ & $54.8\pm1.0$ \\
25 & 23:27:31.33 & -00:42:34.7 & $25.35\pm0.05$ & $3.116$ & $-20.06\pm0.10$ & $41.01\pm1.63$ & $3.47\pm0.14$ & $37.4\pm3.9$ & $8.4\pm1.1$ \\
26 & 23:27:22.89 & -00:41:35.7 & $25.26\pm0.06$ & $3.131$ & $-20.42\pm0.07$ & $35.90\pm2.14$ & $3.08\pm0.18$ & $23.6\pm2.2$ & $6.1\pm2.2$ \\
27 & 23:27:26.97 & -00:41:31.5 & $25.75\pm0.08$ & $3.161$ & $-19.99\pm0.11$ & $10.97\pm1.25$ & $0.96\pm0.11$ & $11.0\pm1.7$ & $6.8\pm2.4$ \\
28 & 23:27:30.20 & -00:42:60.0 & $26.21\pm0.15$ & $3.193$ & $-19.78\pm0.14$ & $2.68\pm0.73$ & $0.24\pm0.07$ & $3.3\pm1.0$ & $6.3\pm9.8$ \\
29 & 23:27:25.03 & -00:43:18.1 & $25.17\pm0.07$ & $3.247$ & $-20.57\pm0.07$ & $1.78\pm0.60$ & $0.17\pm0.06$ & $1.1\pm0.4$ & $1.4\pm3.0$ \\
30 & 23:27:26.09 & -00:41:19.2 & $24.93\pm0.04$ & $3.254$ & $-20.12\pm0.11$ & $3.36\pm1.20$ & $0.32\pm0.11$ & $3.2\pm1.2$ & $8.0\pm5.5$ \\
31 & 23:27:23.96 & -00:44:32.3 & $26.28\pm0.11$ & $3.338$ & $-19.70\pm0.18$ & $2.54\pm0.70$ & $0.25\pm0.07$ & $3.8\pm1.3$ & $5.0\pm3.3$ \\
32 & 23:27:22.68 & -00:44:03.1 & $26.33\pm0.17$ & $3.494$ & $-19.39\pm0.28$ & $6.48\pm1.08$ & $0.72\pm0.12$ & $14.3\pm4.9$ & $6.6\pm2.4$ \\
33 & 23:27:36.40 & -00:41:30.2 & $23.91\pm0.02$ & $3.526$ & $-21.21\pm0.06$ & $18.05\pm2.76$ & $2.05\pm0.31$ & $7.7\pm1.3$ & $3.5\pm2.2$ \\
34 & 23:27:20.52 & -00:42:26.2 & $25.16\pm0.05$ & $3.541$ & $-20.32\pm0.14$ & $12.62\pm1.32$ & $1.45\pm0.15$ & $12.3\pm2.1$ & $6.7\pm3.9$ \\
\hline
\multicolumn{10}{c}{Protocluster candidate of ID15} \\
1 & 16:09:20.62 & +55:33:36.7 & $24.29\pm0.03$ & $2.628$ & $-20.20\pm0.05$ & $825.21\pm12.61$ & $46.35\pm0.71$ & $436.3\pm21.7$ & $3.7\pm5.4$ \\
2 & 16:09:06.79 & +55:30:22.8 & $24.94\pm0.07$ & $2.652$ & $-19.76\pm0.08$ & $59.75\pm4.37$ & $3.43\pm0.25$ & $48.3\pm5.3$ & $0.7\pm2.1$ \\
3 & 16:08:56.00 & +55:31:09.7 & $25.69\pm0.11$ & $2.760$ & $-18.80\pm0.18$ & $8.89\pm3.22$ & $0.56\pm0.20$ & $19.3\pm7.8$ & $10.2\pm8.6$ \\
\hline
\end{tabular}
\end{table*}

\setcounter{table}{1}
\begin{table*}
\caption{\it -- continued}
\label{tab:spec}
\begin{tabular}{cccccccccc}
\hline
No. & R.A. & Decl. & $m_r$ & $z$ & $M_\mathrm{UV}$ & $f_\mathrm{Ly\alpha}$ & $L_\mathrm{Ly\alpha}$ & $EW_0$ & $S_w^a$ \\
 & (J2000) & (J2000) & (mag) & & (mag) & ($10^{-18}\,\mathrm{erg\,s^{-1}\,cm^{-2}}$) & ($10^{42}\,\mathrm{erg\,s^{-1}}$) & (\AA) & (\AA) \\
\hline
4 & 16:08:56.78 & +55:31:36.3 & $23.85\pm0.02$ & $2.810$ & $-21.21\pm0.02$ & $12.21\pm2.89$ & $0.81\pm0.19$ & $3.0\pm0.7$ & $7.5\pm3.3$ \\
5 & 16:09:24.85 & +55:32:50.0 & $23.27\pm0.01$ & $2.899$ & $-21.95\pm0.01$ & $303.47\pm7.62$ & $21.60\pm0.54$ & $40.7\pm1.1$ & $8.7\pm1.3$ \\
6 & 16:09:12.61 & +55:32:17.6 & $25.18\pm0.06$ & $2.909$ & $-20.04\pm0.07$ & $7.68\pm2.32$ & $0.55\pm0.17$ & $6.0\pm1.9$ & $1.1\pm2.3$ \\
7 & 16:08:57.15 & +55:33:48.1 & $25.64\pm0.10$ & $2.924$ & $-19.47\pm0.12$ & $12.60\pm4.36$ & $0.92\pm0.32$ & $16.9\pm6.2$ & $5.4\pm19.5$ \\
8 & 16:09:18.57 & +55:31:42.5 & $24.51\pm0.04$ & $2.931$ & $-20.48\pm0.05$ & $15.84\pm1.92$ & $1.16\pm0.14$ & $8.4\pm1.1$ & $13.3\pm7.4$ \\
9 & 16:09:23.36 & +55:31:43.6 & $24.15\pm0.03$ & $2.940$ & $-21.18\pm0.03$ & $35.42\pm6.78$ & $2.61\pm0.50$ & $10.0\pm1.9$ & $4.4\pm2.6$ \\
10 & 16:09:14.66 & +55:33:10.4 & $25.48\pm0.11$ & $2.957$ & $-19.82\pm0.09$ & $10.00\pm1.65$ & $0.75\pm0.12$ & $10.0\pm1.8$ & $-6.4\pm11.8$ \\
11 & 16:09:13.41 & +55:34:11.2 & $26.26\pm0.17$ & $2.979$ & $-18.57\pm0.31$ & $12.99\pm2.46$ & $0.99\pm0.19$ & $42.0\pm15.9$ & $3.7\pm2.6$ \\
12 & 16:09:03.82 & +55:31:57.9 & $25.07\pm0.07$ & $2.987$ & $-20.24\pm0.06$ & $11.72\pm2.13$ & $0.90\pm0.16$ & $8.2\pm1.6$ & $4.2\pm2.4$ \\
13 & 16:09:16.88 & +55:32:09.0 & $24.10\pm0.02$ & $2.992$ & $-21.16\pm0.03$ & $12.73\pm1.47$ & $0.98\pm0.11$ & $3.8\pm0.4$ & $8.2\pm16.0$ \\
14 & 16:09:22.92 & +55:33:47.9 & $24.97\pm0.06$ & $2.995$ & $-20.28\pm0.06$ & $13.62\pm4.17$ & $1.05\pm0.32$ & $9.2\pm2.9$ & $4.5\pm7.0$ \\
15 & 16:09:19.67 & +55:33:35.0 & $25.36\pm0.08$ & $3.005$ & $-20.14\pm0.07$ & $35.35\pm3.56$ & $2.74\pm0.28$ & $27.4\pm3.3$ & $4.6\pm3.2$ \\
16 & 16:09:19.22 & +55:35:25.2 & $25.57\pm0.10$ & $3.041$ & $-19.49\pm0.13$ & $7.60\pm2.58$ & $0.61\pm0.21$ & $11.1\pm4.0$ & $0.5\pm2.0$ \\
17 & 16:09:11.18 & +55:32:59.5 & $25.91\pm0.11$ & $3.060$ & $-19.48\pm0.16$ & $25.41\pm4.25$ & $2.06\pm0.34$ & $37.6\pm8.6$ & $6.1\pm3.2$ \\
18 & 16:08:50.73 & +55:33:35.0 & $24.15\pm0.03$ & $3.070$ & $-21.50\pm0.02$ & $7.62\pm2.85$ & $0.62\pm0.23$ & $1.8\pm0.7$ & $7.3\pm29.1$ \\
19 & 16:09:09.31 & +55:32:13.3 & $25.56\pm0.08$ & $3.070$ & $-20.15\pm0.07$ & $6.04\pm0.95$ & $0.49\pm0.08$ & $4.9\pm0.8$ & $4.1\pm6.8$ \\
20 & 16:09:07.57 & +55:32:04.9 & $25.49\pm0.09$ & $3.077$ & $-19.99\pm0.08$ & $4.96\pm0.70$ & $0.41\pm0.06$ & $4.7\pm0.8$ & $3.4\pm5.8$ \\
21 & 16:09:15.62 & +55:33:33.8 & $25.99\pm0.13$ & $3.084$ & $-19.44\pm0.14$ & $12.74\pm2.03$ & $1.05\pm0.17$ & $20.1\pm4.2$ & $5.6\pm38.0$ \\
22 & 16:09:02.43 & +55:30:55.8 & $25.14\pm0.08$ & $3.088$ & $-20.09\pm0.08$ & $13.99\pm2.02$ & $1.16\pm0.17$ & $12.1\pm2.0$ & $7.6\pm5.9$ \\
23 & 16:09:26.70 & +55:33:53.8 & $26.40\pm0.20$ & $3.091$ & $-18.69\pm0.26$ & $4.15\pm1.43$ & $0.34\pm0.12$ & $13.1\pm5.7$ & $3.7\pm2.8$ \\
24 & 16:09:09.41 & +55:33:33.8 & $25.19\pm0.07$ & $3.102$ & $-20.15\pm0.08$ & $7.66\pm1.99$ & $0.64\pm0.17$ & $6.3\pm1.7$ & $5.5\pm3.6$ \\
25 & 16:08:56.51 & +55:32:11.9 & $25.76\pm0.12$ & $3.108$ & $-19.62\pm0.12$ & $9.32\pm1.41$ & $0.78\pm0.12$ & $12.6\pm2.4$ & $7.8\pm21.6$ \\
26 & 16:09:06.77 & +55:33:26.5 & $25.33\pm0.08$ & $3.112$ & $-19.89\pm0.10$ & $1.85\pm0.37$ & $0.16\pm0.03$ & $2.0\pm0.4$ & $0.4\pm20.5$ \\
27 & 16:09:10.40 & +55:29:49.6 & $26.20\pm0.17$ & $3.114$ & $-19.33\pm0.19$ & $27.73\pm3.95$ & $2.34\pm0.33$ & $49.2\pm11.6$ & $9.3\pm4.0$ \\
28 & 16:08:51.50 & +55:34:22.1 & $26.15\pm0.13$ & $3.185$ & $-18.74\pm0.28$ & $9.96\pm0.75$ & $0.89\pm0.07$ & $32.3\pm9.7$ & $1.0\pm10.8$ \\
29 & 16:08:59.29 & +55:33:37.9 & $24.94\pm0.05$ & $3.234$ & $-20.53\pm0.06$ & $9.22\pm1.44$ & $0.85\pm0.13$ & $6.0\pm1.0$ & $6.3\pm2.0$ \\
30 & 16:09:12.57 & +55:34:49.7 & $25.21\pm0.06$ & $3.261$ & $-20.35\pm0.09$ & $20.18\pm3.05$ & $1.91\pm0.29$ & $15.6\pm2.7$ & $3.4\pm6.5$ \\
31 & 16:08:56.52 & +55:32:25.0 & $25.67\pm0.11$ & $3.262$ & $-20.18\pm0.09$ & $7.24\pm0.97$ & $0.68\pm0.09$ & $6.6\pm1.0$ & $5.6\pm7.6$ \\
32 & 16:09:04.72 & +55:32:46.6 & $26.70\pm0.24$ & $3.271$ & $-19.05\pm0.24$ & $7.86\pm1.86$ & $0.75\pm0.18$ & $20.3\pm6.9$ & $5.4\pm25.8$ \\
33 & 16:08:59.09 & +55:33:27.6 & $25.62\pm0.09$ & $3.277$ & $-20.09\pm0.10$ & $13.84\pm3.78$ & $1.32\pm0.36$ & $13.8\pm4.0$ & $0.9\pm3.1$ \\
34 & 16:08:57.30 & +55:33:20.7 & $26.34\pm0.16$ & $3.361$ & $-19.44\pm0.19$ & $7.59\pm0.79$ & $0.77\pm0.08$ & $14.6\pm3.2$ & $1.6\pm4.6$ \\
35 & 16:08:51.11 & +55:31:50.5 & $26.98\pm0.28$ & $3.386$ & $-18.80\pm0.34$ & $14.79\pm2.03$ & $1.53\pm0.21$ & $52.6\pm20.4$ & $6.2\pm2.5$ \\
36 & 16:09:26.02 & +55:34:06.3 & $26.43\pm0.21$ & $3.390$ & $-18.71\pm0.36$ & $14.89\pm2.66$ & $1.54\pm0.28$ & $57.7\pm25.1$ & $4.7\pm2.0$ \\
37 & 16:09:04.30 & +55:33:50.3 & $24.30\pm0.04$ & $3.422$ & $-21.51\pm0.03$ & $13.87\pm3.04$ & $1.47\pm0.32$ & $4.2\pm0.9$ & $4.4\pm2.7$ \\
\hline
\multicolumn{10}{l}{$^a$ Weighted skewness of the emission line profile, as defined by Equation 2 in \citet{kashikawa06}.}
\end{tabular}
\end{table*}

\section{Results} \label{sec:res}
\subsection{Protocluster confirmation}
\begin{figure}
\begin{center}
\includegraphics[width=0.7\hsize]{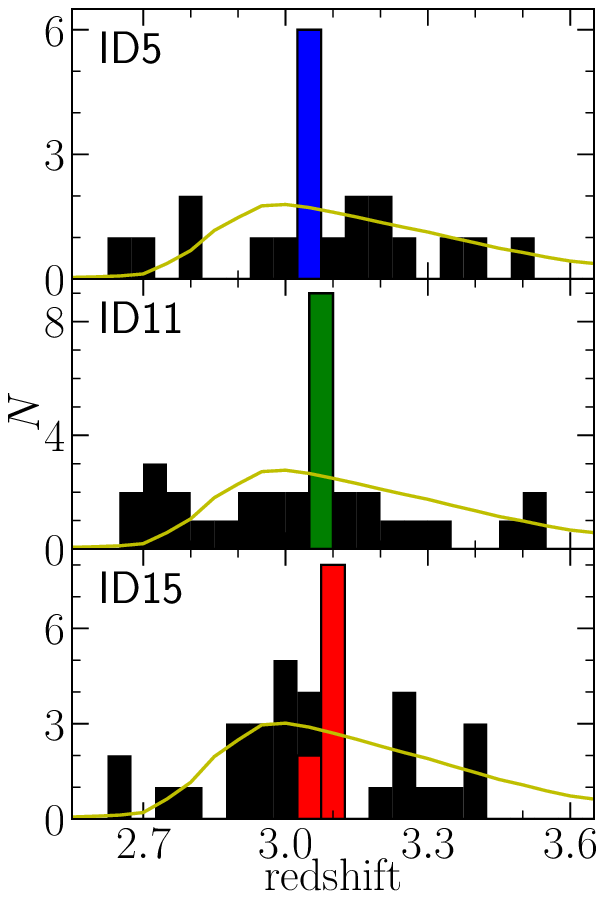}
\caption{Redshift distribution of spectroscopically-confirmed $U$-dropout galaxies in the three target regions of
    protocluster candidates.
    The bin size is $\Delta z=0.05$.
    The yellow lines indicate the redshift selection function of $U$-dropout galaxies.
    The protocluster candidates of ID5, ID11 and ID15 exhibit significant redshift spikes, which are shown by blue, green and
    red bars, respectively.
    Note that the ten galaxies spreading over two red bins in the bottom panel are clustering within
    $\Delta z=0.04$.}
\label{fig:zhist}
\end{center}
\end{figure}
Protoclusters are defined to be the progenitors of local galaxy clusters, so they would not yet be a single
virialised halo and would typically be composed of multiple independent halos at high redshifts \citep[e.g.,][]{chiang17}.
In such an early phase of cluster formation, the morphology of protoclusters, or the spatial distribution of
member galaxies, tends to be more aspherical and filamentary compared with that of local galaxy clusters
\citep{lovell18}.
Thus, both descendant halo mass and viewing angle are crucial parameters to predict the projected size of
protoclusters on the sky, though either parameters could not be controlled in an actual observation.
Typically, protocluster galaxies at $z\sim3$ are expected to extend over a sphere of $R\sim14\,\mathrm{cMpc}$
radius, corresponding to $R_\mathrm{sky}\sim7\,\mathrm{arcmin}$ and $R_z\sim0.015$
\citep[cf. Figure~4 of ][]{toshikawa16}.
As protocluster morphology may differ significantly from spherical symmetry, we define that protocluster members are clustering within
an end-to-end size of $\Delta z=0.05$ and show a statistically significant excess of galaxy number within the
redshift bin.
It should be noted that, as shown in Figure~\ref{fig:cntr}, the size of $>4\sigma$ overdense regions appears to be only a few arcminutes;
however, the outskirts of protoclusters cannot be accurately traced by the surface overdensity of dropout galaxies due to projection effects.
Actual extent of protoclusters should be much larger than regions outlined by $2\sigma$ or $3\sigma$ overdensity contours.
Since we can identify only the tips of protoclusters by the surface overdensity of dropout galaxies due to projection effects, protocluster members are expected to be distributed even in surrounding regions with $\lesssim2\sigma$ overdensity.
We do not consider spatial position for selecting protocluster members from our spectroscopic sample as the FoV of our
spectroscopy ($\sim3\,\mathrm{arcmin}$ radius) is smaller than the typical extent of protoclusters.

Figure~\ref{fig:zhist} presents the redshift distribution of spectroscopically-confirmed galaxies in each target region and the redshift selection window of $U$-dropout galaxies.
The redshift window is estimated by the same procedure as in \citet{toshikawa16} but using the spectral synthesis code of CIGALE \citep{burgarella05,noll09,boquien19}.
There are clear peaks including six, nine, and ten galaxies within $\Delta z<0.05$ at $z=3.05$, $z=3.07$, and
$z=3.09$, in the protocluster candidates of ID5, ID11, and ID15, respectively.
These numbers within a bin of $\Delta z=0.05$ correspond to $\sim3.5$ times higher than the expected from a random distribution.
If expected number is calculated according to the exact redshift size ($\Delta z=0.027\mathrm{-}0.043$) of individual protoclusters, the number of protocluster members correspond to a $\sim4\mathrm{-}6$ times higher.
In order to evaluate the statistical significance of this clustering, we have estimated a probability to reproduce
such tight clustering from random distributions.
A mock redshift distribution is reproduced by randomly selecting the same number of redshifts as that of
spectroscopically-confirmed galaxies ($N=22$, 34, or 37 in the ID5, ID11, or ID15) from the redshift selection
function of $U$-dropout galaxies; then, it is checked whether there is a comparable or more clustering galaxy
group.
After repeating this procedure 1,000 times, random distributions are found to be able to mimic the same level of
clustering as in the protocluster candidates of ID5, ID11, and ID15 with a probability of 1.6\%, 0.3\%, and 0.5\%,
respectively.
These probabilities are small enough to conclude that such clustering groups cannot be caused by chance, and we
have spectroscopically confirmed three protoclusters.

The ID5, ID11, and ID15 protoclusters are composed of at least six (No.7-12), nine (No.16-24), and ten (No.18-27)
galaxies within the redshift range of $\Delta z=0.030$, $\Delta z=0.027$, and $\Delta z=0.043$, respectively.
It should be noted that the number of protocluster members corresponds to $\sim25\%$ of all spectroscopically-confirmed galaxies though the surface overdensity exhibits about twice higher ($\sim50\%$ excess).
As described in \citet{toshikawa24}, the completeness of our protocluster sample is low ($\sim10\%$) due to projection effects.
Protocluster candidates would be preferentially selected if their surface overdensity is boosted by foreground/background galaxies.
As for the success rate of protocluster confirmation in our follow-up spectroscopy,
according to the model comparison of \citet{toshikawa24}, $\sim90\%$ of protocluster candidates are expected to include
genuine protoclusters; at least $\sim75\%$ of candidates can exhibit a clustering signal strong enough to be
confirmed by actual observations.
Considering small-number statistics \citep[e.g.,][]{gehrels86}, it is still reasonable that all three targets are genuine protoclusters.
However, precisely speaking, we have just confirmed statistically significant clustering, and it is difficult to accurately relate clustering strength at high redshifts to descendant halo mass at $z=0$ based on the currently available dataset.
Our sample of confirmed protoclusters may be contaminated by the progenitors of smaller systems like galaxy groups
instead of rich clusters, so that the apparent success rate of protocluster confirmation might be enhanced.
Irrespective of the absolute scale/mass, our sample of protoclusters pinpoints higher-density environments at
$z\sim3$, allowing us to investigate environmental effects on galaxy evolution in the high-redshift Universe.

In the following sections, we will deduce physical properties from the confirmed protocluster members.
We acknowledge the confirmed members would be just a part of the protoclusters.
At first, our method to identify protoclusters is completely dependent on a specific galaxy population of dropout
galaxies having Ly$\alpha$ emissions, resulting in a bias towards young, star-forming galaxy populations.
In addition, the typical spatial size of protoclusters is about two times larger than that of the observed area.
Even within the FoV of our follow-up spectroscopy, we have observed only about a third of photometrically-selected,
$U$-dropout galaxies due to the limitation of slit allocation.
Although this observational bias and incompleteness make it difficult to reconstruct a full picture of cluster
formation or galaxy evolution in high-density environments, we utilise the sample of foreground/background galaxies
as field counterparts because they are identified from the same method/dataset as the protocluster members themselves.
Such counterpart selection hence reduces the effects of bias/incompleteness and enables highlighting the physical nature of
protoclusters.

\subsection{Ly$\alpha$ and UV properties}
\begin{figure}
\includegraphics[width=\hsize]{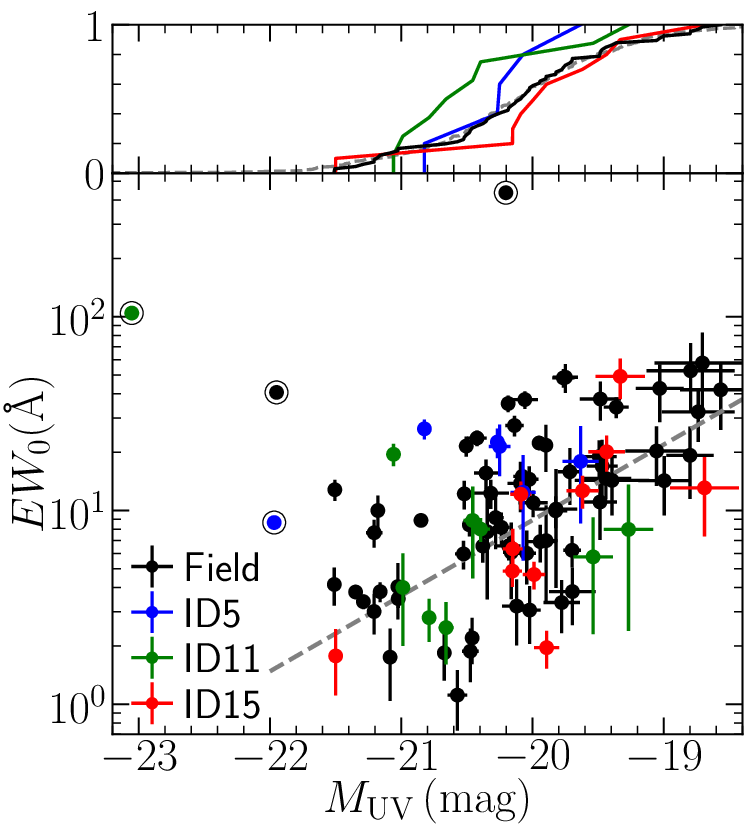}
\caption{Bottom: Relation between $M_\mathrm{UV}$ and $EW_0$ of spectroscopically-confirmed $U$-dropout galaxies. The gray dashed line is the best-fitting linear relation between $M_\mathrm{UV}$ and $\log(EW_0)$. The member galaxies of ID5, ID11, and ID15 protoclusters and field galaxies are indicated by the blue, green, red, and black points, respectively. $U$-dropout galaxies showing AGN signature are marked by the open circles.
Top: The cumulative $M_\mathrm{UV}$ distributions of protoclusters (coloured) and field galaxies (black) without AGNs. That of spectroscopically-observed galaxies is shown by the dashed gray line, which overlaps that of field galaxies with Ly$\alpha$ emissions (solid black line). This indicates that the observational bias of Ly$\alpha$ detection does not alter the $M_\mathrm{UV}$ distribution of spectroscopic targets.}
\label{fig:MuvEW}
\end{figure}
\begin{figure}
\includegraphics[width=\hsize]{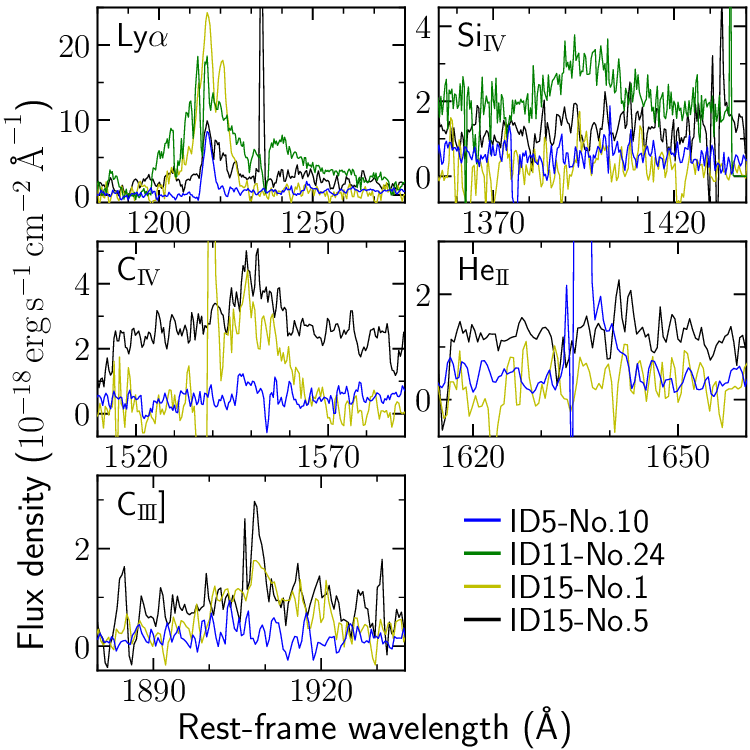}
\caption{Spectra of four AGNs focusing on the emissions of Ly$\alpha$ (upper left), Si{\sc iv} (upper right),
    C{\sc iv} (middle left), He{\sc ii} (middle right), and C{\sc iii}] (bottom).}
\label{fig:AGN}
\end{figure}
As both of UV continuum and Ly$\alpha$ emission are indicators of star-formation rate (SFR), $M_\mathrm{UV}$ and $EW_0$ are
correlated. Our combined sample of member galaxies in the protoclusters shows a consistent $M_\mathrm{UV}$-$EW_0$
relation with that of field galaxies as shown in Figure~\ref{fig:MuvEW}.
A few galaxies, marked by the open circles in Figure~\ref{fig:MuvEW}, are located above this relation.
All of them are found to be AGNs by detecting high-ionisation emission lines (Si{\sc iv}, C{\sc iv}, He{\sc ii},
or C{\sc iii}]), as well as broad Ly$\alpha$ emission line (Figure~\ref{fig:AGN}).
The fraction of AGNs among $U$-dropout galaxies seems to be higher in the protoclusters (2/25 or 8.0\%) than in the field (2/68 or 2.9\%), though it is only a $1.5\sigma$ significance difference according to the binomial distribution. 
It should be noted that the AGN in the ID5 protocluster (ID5-No.10) has weak C{\sc iv}, C{\sc iii}] emissions but strong He{\sc ii} emission, while typical
AGNs have the line ratio of $\mathrm{C_{I\hspace{-0.6pt}V}/He_{I\hspace{-0.6pt}I}}>1$.
Such a smaller line ratio of C{\sc iv}/He{\sc ii} could result from higher metallicity \citep{nagao06}.

\begin{figure}
\includegraphics[width=\hsize]{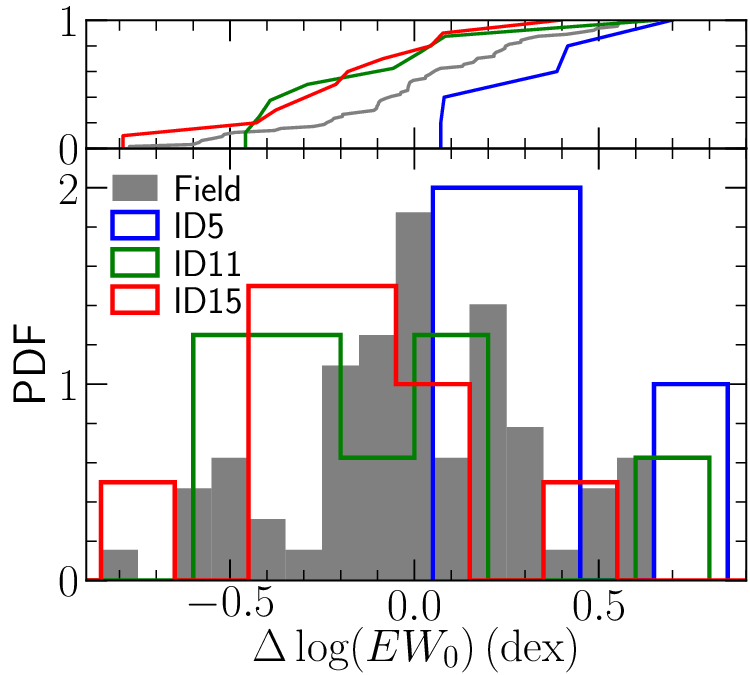}
\caption{Distribution of the deviation of $EW_0$ from the average $M_\mathrm{UV}$-$EW_0$ relation (lower panel).
    The average relation is derived by fitting a linear function to all spectroscopically-confirmed galaxies except
    for AGNs.
    The gray histogram indicates field galaxies, and the blue/green/red histograms indicate the ID5/ID11/ID15
    protoclusters, respectively.
    The upper panel shows the cumulative distribution functions of the deviation.}
\label{fig:EW}
\end{figure}

Digging deeper, we note that individual protoclusters exhibit distinguishing characteristics from each other.
Given fixed $M_\mathrm{UV}$, the member galaxies of the ID5 protocluster tend to have higher $EW_0$, while the
Ly$\alpha$ emissions from the ID11 and ID15 protoclusters are fainter.
Discarding AGNs, the relation between $M_\mathrm{UV}$ and $\log(EW_0)$ of field galaxies can be well fitted by a linear function of $\log(EW_0)=0.39\times M_\mathrm{UV}+8.75$.
Although this relation is determined by combining all three regions, the relation in each region is also found to be consistent within $1\sigma$ uncertainty.
This consistency also suggests that there is less observational bias among the three target regions.
The distribution of deviations from this average relation are shown in Figure~\ref{fig:EW}, making it more apparent that the Ly$\alpha$ properties of the ID5 protocluster is significantly higher than that of either field galaxies or the other protoclusters.
The statistical significance of these differences is found to be $p<0.05$ according to the Kolmogorov–Smirnov (KS) test.
Although the ID11 and ID15 protoclusters have smaller $\Delta\log(EW_0)$ on average than that of field galaxies, their statistical significance is marginal.

When we focus on individual galaxies in the protoclusters, we can notice some interesting features.
The $M_\mathrm{UV}$ distribution of the ID15 protocluster members is slightly different from that of field
galaxies; especially, there is a large gap between the first and second UV brightest galaxies in the ID15
protocluster.
Its $M_\mathrm{UV}$ gap is $1.3\,\mathrm{mag}$, which may resemble a context for brightest cluster galaxies (BCGs) in the local Universe \citep[cf.][]{ito19}.
To evaluate whether such large gap may be attributed to the incomplete sampling of protocluster members, or small number statistics, we carry out the following exercise. The ID15 protocluster counts 10 confirmed members. We repeatedly create identically sized samples of objects randomly drawn from the field galaxy population. Then, we checked for each such mock sample whether the largest magnitude gap exceeded the observed one (1.3 mag).
The probability to reproduce the observed $M_\mathrm{UV}$ gap in the ID15 protocluster is thus found to be less than 0.5\%,
suggesting that the brightest galaxy in ID15 protocluster is exceptional.
We further note that there are also large gaps between the 1st and 2nd brightest members in the ID5 and ID11
protoclusters, though the brightest ones are AGNs.
No clear signature of AGN activity is found in the brightest member of protocluster ID15, which features a small
Ly$\alpha$ flux.
While the enhanced abundance of UV bright galaxies in the spectroscopic sample of the three protoclusters is consistent with what we have concluded based on photometric datasets \citep{ito20,toshikawa24}, the additional spectroscopy performed in this study reveals that the presence of AGNs contributes to such bright population.
The overall $M_\mathrm{UV}$ distribution without AGNs is not significantly different between protoclusters and field galaxies as shown in the top panel of Figure~\ref{fig:EW}.

\subsection{Physical properties by SED fitting \label{sec:SEDfit}}
\begin{figure}k
\includegraphics[width=\hsize]{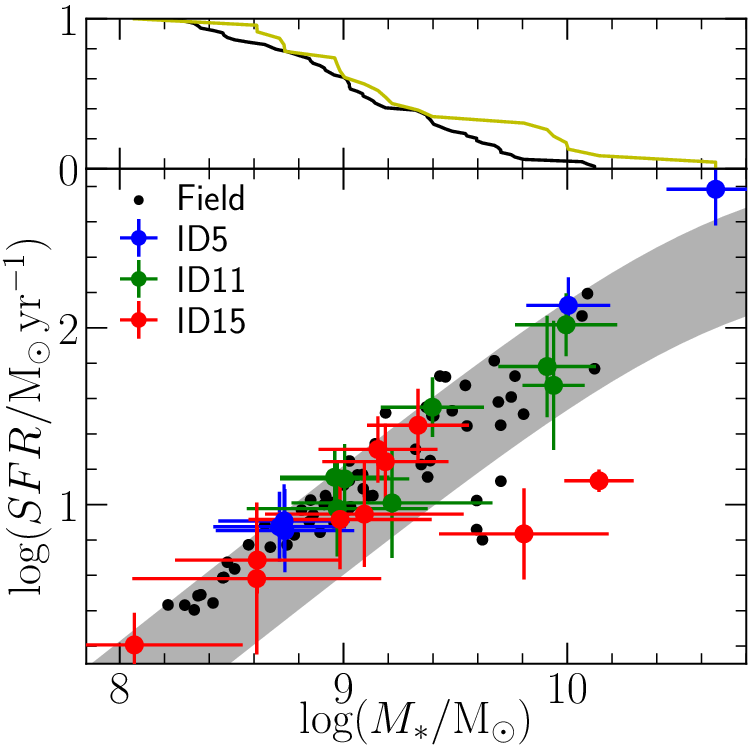}
\caption{The relation between $M_*$ and SFR derived by SED fitting (see Section~\ref{sec:SEDfit} for details).
    The black, blue, green and red points represent field, ID5, ID11 and ID15 protocluster galaxies, respectively.
    The error bars are not plotted for field galaxies for clear visualisation, but the size of errors is similar to
    that for protocluster members as they are estimated from the same dataset.
    The gray region indicates the main-sequence of star-forming galaxies in general fields with
    $\pm0.3\,\mathrm{dex}$ from \citet{popesso23}.
    The top panel represents the cumulative distribution function of stellar mass.
    The yellow and black lines are for protocluster and field galaxies, respectively.
    Protocluster galaxies show an excess at the massive end.}
\label{fig:MS}
\end{figure}
Besides the optical imaging from CFHT/$U$-band to HSC/$y$-band, near-infrared (NIR) images are available in
the DUD layer of the HSC-SSP.
$J$-, $H$-, and $K$-band imaging were taken by the Deep UKIRT NEar-Infrared Steward Survey (DUNES$^2$).
Almost the entire area of the DUD layer is covered by deep NIR images thanks to the $\sim260\,\mathrm{hours}$ of observing time with UKIRT/WFCAM allocated to DUNES$^2$ and additional archival data (e.g., UltraVISTA and
UKIDSS).
Furthermore, Spitzer/IRAC observations were carried out by the Spitzer Coverage of HSC-Deep with IRAC for Z studies
\citep[SHIRAZ;][]{annunziatella23}.
Combined with archival IRAC data, SHIRAZ produced deep IRAC ch1 \& ch2 images over $\sim17\,\mathrm{deg^2}$
area of the HSC-SSP DUD layer.
All three target regions of this study are covered by this multi-wavelength dataset from CFHT/$U$-band to IRAC/ch2.

We are able to estimate the stellar population of high-redshift galaxies by spectral energy distribution (SED) fitting with multi-wavelength photometry.
For an accurate estimate, it is critical to constrain fluxes at longer wavelengths than the Balmer/4000{\AA} break, corresponding to $K$-band or IRAC.
The three target regions have a $2\sigma$ depth of $\sim24.5\,\mathrm{mag}$ at $K$-band.
The $2\sigma$ depth of IRAC/ch1(ch2) is $\sim24.7\,(24.4)\,\mathrm{mag}$ and $\sim23.6\,(23.4)\,\mathrm{mag}$ in the regions of the ID5 and ID11/ID15 protoclusters, respectively.
As the ID5 protocluster is located in COSMOS, its depth at Spitzer/IRAC is $\sim1\,\mathrm{mag}$ deeper.
Of 89 spectroscopically-confirmed $U$-dropout galaxies without four AGNs, 41 are detected by either $K$-band, IRAC ch1, or ch2 at $S/N>2.0$.
Thanks to the deep NIR dataset, even non-detection ($S/N<2$) at NIR wavelengths can be useful to constrain the physical properties of galaxies by including upper limits in SED fitting.
We used the SED fitting code CIGALE \citep{burgarella05,noll09,boquien19}, which generates various types of SEDs by assuming, for example, a star formation history, stellar population synthesis model, metallicity, and dust attenuation.
The set of broad-band fluxes computed by integrating the various model SEDs are compared to the observed fluxes to determine a best-fitting physical quantities (e.g., stellar mass, SFR, or age) by $\chi^2$ analysis.

Specifically, we have employed the following models/modules of CIGALE: 1) delayed SFH, $\mathrm{SFR}(t)\propto t\exp(-t/\tau)$,
where $t$ ($=25\,\mathrm{Myr}\mathrm{-}2.0\,\mathrm{Gyr}$ with $\Delta\log t=0.1$) and $\tau$
($=10\,\mathrm{Myr}\mathrm{-}10\,\mathrm{Gyr}$ with $\Delta\log \tau=0.1$) are the age since the onset of star
formation and the time at the peak of star formation, 2) the population synthesis model of \citet{bc03} with
\citet{chabrier03} IMF and stellar metallicity of $0.2Z_{\sun}$ or $0.4Z_{\sun}$, and 3) the dust attenuation law
of \citet{calzetti00} with the parameter range of $E(B-V)=0.0\mathrm{-}1.0$ ($\Delta E(B-V)=0.05$).
Redshifts are fixed to the spectroscopic values, which are determined by the peak wavelength of Ly$\alpha$ emission.
CIGALE uses the IGM transmission prescription of \citet{meiksin06} and cannot set its optical depth as a free
parameter.
However, \citet{thomas17} found a large dispersion of IGM transmission depending on line-of-sight directions; thus,
we do not use $U$-band photometry in our SED fitting since it is heavily affected by IGM absorption.

Rather than adopting the physical parameter values at the minimum $\chi^2$, we use physical parameters corresponding to a likelihood-weighted mean and standard deviation as the fitting
results of value and error, where likelihood $\mathcal{L}$ is expressed as $\mathcal{L}\propto\exp(-\chi^2/2)$. This approach can alleviate intrinsic degeneracy between physical properties or the artificial minimisation of $\chi^2$ due to the small number of data points. Any observed fluxes with $S/N < 2.0$ are replaced by $2\sigma$ upper limits, which deteriorate the $\chi^2$ value only if model fluxes exceed the upper limits. Following these steps, CIGALE produces various model SEDs and calculates $\chi^2$ for each grid point. We note that 97\% (100\%) of our sample have a reduced $\chi^2$ of $\chi_r^2<3\,(5)$ at their best-fitting models. As this $\chi_r^2$ distribution is well matched to the normal distribution, our SED fitting works well for all galaxies without technical problems, such as over-fitting.

The physical properties of spectroscopically-confirmed $U$-dropout galaxies, except for AGNs, are derived following the above method. Figure~\ref{fig:MS} shows the resulting relation between stellar mass ($M_*$) and $SFR$.
The fraction of galaxies which are detected at wavelengths longer than rest-frame 4000{\AA} ($K$-band or IRAC) is
100\% and 74\% at $\log(M_*/\mathrm{M_\odot})>9.5$ and $\log(M_*/\mathrm{M_\odot})>9.0$, respectively.
There is no significant difference between protocluster and field galaxies in terms of the overall trends of these
properties.
Most protocluster members follow the main sequence of star-forming galaxies \citep{popesso23}, as expected by
our sample selection of dropout galaxies.
However, the $M_*$ distribution of protocluster galaxies seems to be different from that of field galaxies
(Figure~\ref{fig:MS}).
Protocluster galaxies are biased towards the massive end, and this feature becomes the most prominent by the threshold
of $M_*>10^{9.8}\,\mathrm{M_\odot}$, resulting in 30\% ($7/23$) of protocluster galaxies compared with 6.3\%
($4/64$) in the field.
According to the binomial distribution, the fraction of massive galaxies in protoclusters is significantly higher
than that of field galaxies with $>99\%$ confidence.
Although the statistical significance depends on the $M_*$ threshold considered to identify massive galaxies, the confidence level remains higher than $90\%$ for any $M_*$ threshold $>10^{9.6}\,\mathrm{M_\odot}$.
It should be noted that the typical uncertainty of $M_*$ estimates by SED fitting is $0.3\,(0.2)\,\mathrm{dex}$ for the
full (massive) sample; thus the chance of a false positive result due to statistical fluctuations in $M_*$ estimates is
negligible.

In addition to the this common feature, we see a sign of uniqueness in each individual protocluster.
For example, the ID5 protocluster seems to be bimodal, with no members at
$M_*\sim10^9\mathrm{-}10^{10}\,\mathrm{M_\odot}$.
At the massive end, member galaxies in the ID5 protocluster are still actively forming stars while those in the
ID15 protocluster are located below the main sequence, suggesting that some members have the signs of a quenching phase.
As for the ID11 protocluster, its massive galaxies follow the main sequence of coeval field galaxies.
The small number of member galaxies in a protocluster makes it difficult to statistically validate the reliability
of these results, but differences between protoclusters may reveal their developmental phase of
cluster formation.

\section{Discussion} \label{sec:dis}

\subsection{Developmental phase of protoclusters} \label{sec:MS}
\begin{figure*}
\includegraphics[width=0.80\hsize]{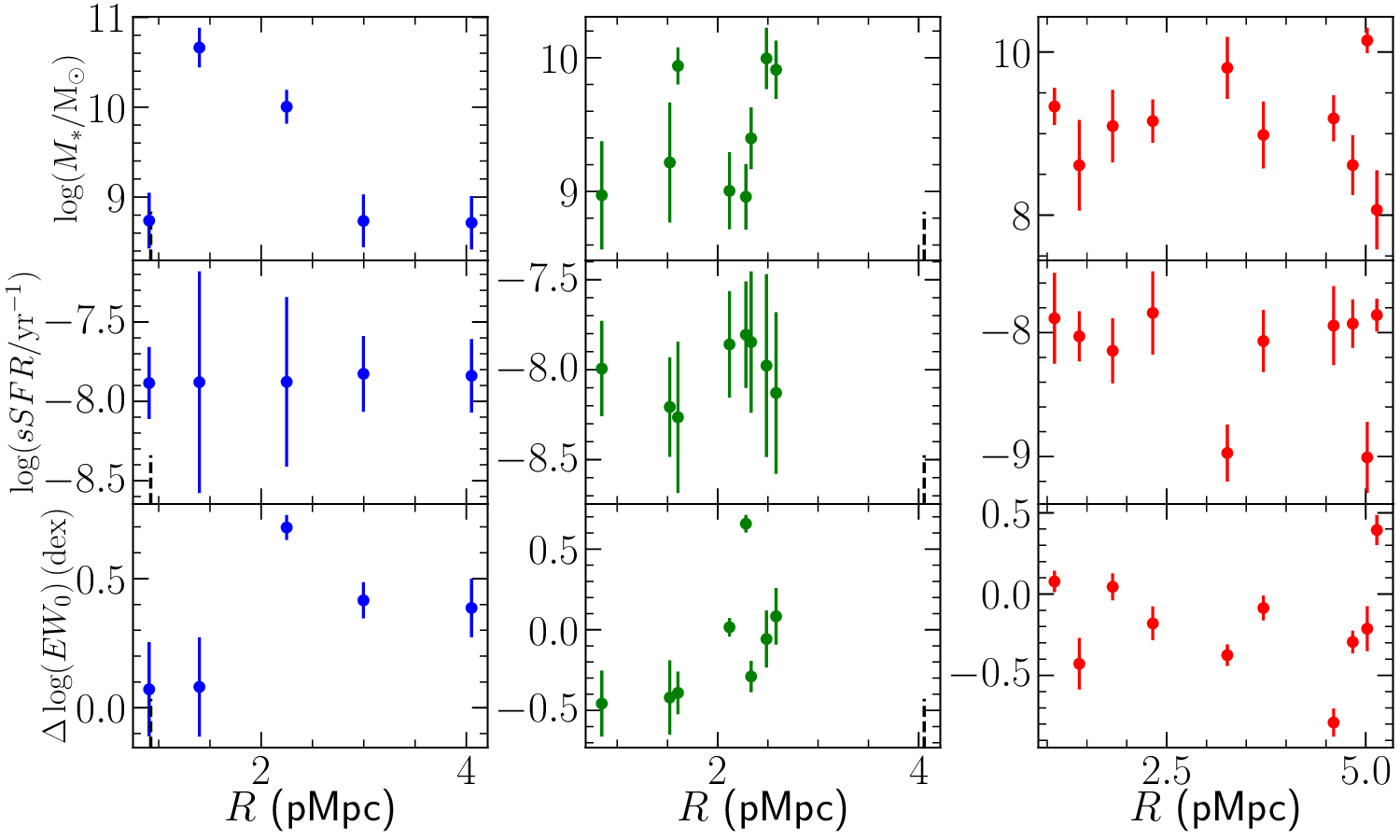}
\caption{Radial dependence of galaxy properties of stellar mass (top), $sSFR$ (middle), and Ly$\alpha$ strength
    (bottom).
    The results for the ID5, ID11, and ID15 protoclusters are shown from left to right panels.
    Due to the limited photometric data, the host galaxy properties of AGNs found in the ID5 and ID11 protoclusters
    cannot be derived by our SED fitting; thus, just their radial positions are indicated by the vertical dashed
    lines near the $x$-axes of the left and middle panels.}
\label{fig:galpropR}
\end{figure*}
We have not only discovered three protoclusters at $z\sim3$, but importantly identified them from the same dataset and using identical methods.
In spite of our uniform way to find protoclusters, they exhibit distinct properties, such as Ly$\alpha$ strength
(Figure~\ref{fig:EW}) and star-forming activity of massive protocluster members (Figure~\ref{fig:MS}). We thus expect these variations to be inherent to the protocluster nature rather than just an imprint of observational
bias.
Theoretical models predict that protoclusters are dominated by young, star-forming galaxies in their earlier
developmental phase; then, quiescent galaxies gradually come into existence \citep[e.g.,][]{contini16,chiang17}.
In this context of cluster formation, based on Figure~\ref{fig:MS}, the ID15 protocluster would be a relatively
mature system since its massive members are below the main sequence of star-forming galaxies.
On the other hand, the ID5 and ID11 protoclusters would represent earlier and intermediate phases, respectively, based on
the star-forming activity of their massive galaxies.
On top of the star-forming activity of massive galaxies, the ID15 protocluster uniquely has the large gap of UV luminosity between the first and second brightest member galaxies.
\citet{sawicki20} found a similar luminosity gap between BCG progenitors and their companions at $z\sim1.6$ and predicted that BCG progenitors have experienced major mergers at even higher redshifts, which would cause the rapid growth of most massive galaxies while less-massive companions are not significantly affected.
If this is the case, the presence of a large gap also suggests that the ID15 protocluster is a relatively mature system.
\begin{figure}
\includegraphics[width=\hsize]{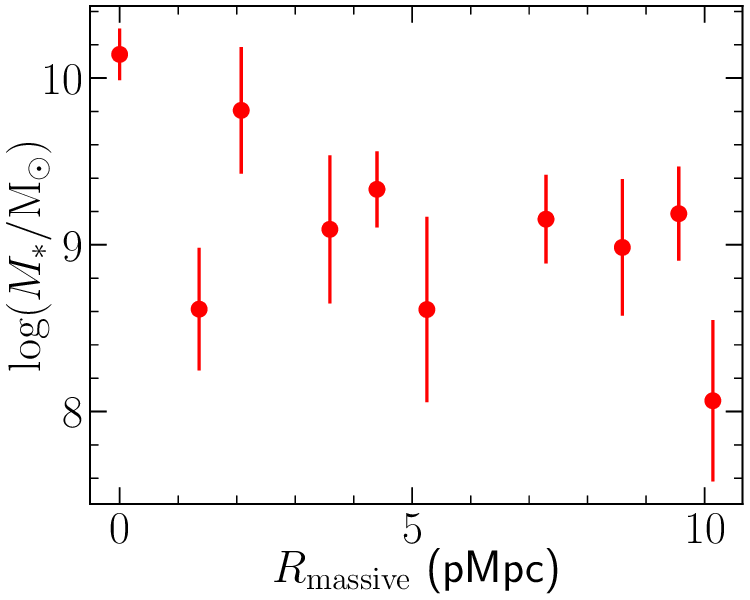}
\caption{Radial dependence of stellar mass in the ID15 protocluster.
    This is similar to the bottom right panel of Figure~\ref{fig:galpropR}, but the point of origin for radial
    distance is determined by the position of the most massive member galaxy instead of the geometric centre.}
\label{fig:MstarRmass}
\end{figure}

The 3D distribution of member galaxies, or the internal structure of a protocluster, is also a useful parameter
to judge the developmental phase.
The accretion of galaxies or mass through the cosmic web is predicted to make protoclusters elongated or filamentary in
their earlier phase; subsequently, protoclusters become dynamically relaxed, with member galaxies that are more spherically
distributed \citep[e.g.,][]{lovell18}.
In this study, the outskirts of protoclusters may be out of the FoV of follow-up spectroscopy, and only a part of
protocluster members could be identified.
Such observational limitations hamper the measurement of protocluster morphology.
Therefore, we have simply calculated the centers of protoclusters by the mean of the sky positions (R.A. and Decl.)
and redshifts of protocluster members and investigated the radial dependence of galaxy properties.
Galaxy evolution is predicted to preferentially progresses from the inner, or denser, regions of (proto)clusters
\citep[e.g.,][]{burg15}, which naively would result in a central concentration of massive, evolved galaxies.
However, in our sample such galaxy properties (e.g., $M_*$ and $sSFR$) do not correlate with radial distance (Figure~\ref{fig:galpropR}).
Furthermore, the quenching galaxies in the ID15 protocluster are located in the outer region.

AGNs are another unique galaxy population, which may differentiate protoclusters from each other because AGN activity is expected to be triggered by galaxy merger or gas accretion.
We have identified AGNs in the ID5 and ID11 protoclusters, which are expected to be in earlier developmental stages than the ID15 protocluster based on the star-forming activity of member galaxies as discussed above.
This result is consistent with \citet{toshikawa20}, in which two protoclusters at $z\sim4$ are observed and an
AGN is found in the one with an earlier developmental phase.
Both higher and lower AGN fractions in protoclusters at $z\sim2\mathrm{-}3$ are reported compared with coeval
field galaxies \citep{lehmer09,macuga19,polletta21}.
Interestingly, \citet{gatica24} found that the AGN fractions of protocluster candidates selected by Planck compact
sources significantly depend on whether they feature a red SPIRE colour or not.
This subsample selection is empirically defined, and may help to divide protoclusters according to their
physical nature, such as developmental phase or maturity.
It should be noted that, even between the two samples of the ID5 and ID11 protoclusters, we can see a large
difference in AGN location: the AGN in ID5 resides in the innermost regions while the one in ID11 inhabits the outermost regions (see the vertical dashed lines in Figure~\ref{fig:galpropR}).
Galaxy mergers are predicted to act as triggers of AGN activity by both theoretical and observational studies
\citep[e.g.,][]{steinborn18,li23}; subsequently, AGN feedback can halt the star-forming activity of host galaxies.
If this is the case, the difference of AGN location between ID5 and ID11 could reflect the
inside-out growth of galaxy clusters.
The correlation between protoclusters and AGNs is still under debate as a lot of conflicting results are reported
\citep[e.g.,][]{uchiyama20,vito20,chapman26}.
Such inconsistency may also result from the dependence of AGN incidence on the developmental stage of protoclusters.

Although we have found that each protocluster possesses distinctive characteristics of galaxy properties,
there is no significant correlation between such variation and protocluster structure, or radial distance.
It should be noted that the definition of a protocluster centre is not trivial, especially in actual observations.
Due to the protocluster's nature as a far from dynamically-relaxed system, the position of the geometric
centre, the centre of the mass, and the peak of galaxy number density can be shifted with respect to each other
\citep[e.g.,][]{baxter25}.
Similarly, the most massive galaxies are not always located at the centre of their host protocluster at high
redshifts, and \citet{contini16} predicted a typical separation of $\sim2\,\mathrm{pMpc}$ at $z\sim3$.
The estimates of a protocluster's centre, size, and morphology further depend on the galaxy population used to trace the underlying
structures \citep[e.g.,][]{hough20}.
Such observational bias and incompleteness could dilute the signal of any radially dependent galaxy properties.
Therefore, we have also investigated the radial dependence of stellar mass by assuming the position of the most massive galaxy
as a centre.
As shown in Figure~\ref{fig:galpropR}, the most massive galaxy in the ID15 protocluster is $\sim5\,\mathrm{pMpc}$
away from the geometric centre of the protocluster while those in the ID5 and ID11 protoclusters are located near their geometric centre ($R\lesssim2\,\mathrm{pMpc}$).
Thus, changing the point of origin from a geometric centre to the position of the most massive galaxy does not
have a significant impact on the ID5 and ID11 protoclusters, but for the ID15 protocluster leads to a marginal
trend between stellar mass and distance from the most massive galaxy (Figure~\ref{fig:MstarRmass}). 
More massive galaxies tend to be closer to the most massive galaxy, implying that the ID15 protocluster is forming
its core by assembling mass and galaxies.

These observed features in the three protoclusters can be attributed to the difference of the developmental stages of galaxy clusters.
However, this study relies on Ly$\alpha$-emitting dropout galaxies, which would be just a small part of galaxy populations, especially young star-forming galaxies.
Completely quenched galaxies, which may be a sign of mature protoclusters, cannot be identified in this study; similarly, starburst galaxies involving heavy dust-obscuration are also dropped from sample selection.
Actually, \citet{toshikawa25} discovered a protocluster at $z=3.70$ in a consistent manner with this study, and rich multi-wavelength dataset enabled to find that massive galaxies near its centre and sub-mm galaxies on the outskirts.
In addition, \citet{cooke13} confirmed that even the photometric sample of dropout galaxies can be composed of multiple galaxy populations.
They divided dropout galaxies at $z\sim3$ into two subgroups representing Ly$\alpha$ emitters and absorbers according to photometric properties and statistically investigated their spatial distribution and environment; then, a fraction of Ly$\alpha$-emitting dropout galaxies are considered to be located on the outskirts of group-like halos, or shell-like structures.
This tendency would make this study less sensitive to galaxies in the core, where evolved galaxies may preferentially exist.
Galaxy population studies were carried out on several protoclusters at $z\sim2\mathrm{-}3$.
\citet{kuiper10} confirmed a significant radial dependence of galaxy properties in a $z\sim3.1$ protocluster: massive and highly star-forming galaxies, traced by Balmer break galaxies or [O{\sc iii}] emitters, are concentrated at the centre, while young, less massive galaxies, like LAEs, are on the outskirts or widely distributed.
This protocluster would be representative of the inside-out growth of galaxy clusters.
On the other hand, \citet{shi19,shi25} found that overdensity peaks of LBGs, LAEs, or [O{\sc iii}] emitters are spatially offset in protoclusters at $z\sim3$; especially, star-forming and quiescent galaxies show a strong anti-correlation.
According to the hierarchical structure formation model, galaxy clusters are not formed only by spherically accreting mass or galaxies but also by merging group-scale halos.
Each subgroup in a protocluster is dominated by a distinct galaxy population possibly due to their different formation epochs or halo assembly bias.
Such stochastic, anisotropic processes could result in complex distribution or spatial segregation depending on galaxy populations in protoclusters.
As clearly presented by these previous works, it is essential for revealing cluster formation to evaluate the range of diversity of galaxy populations and their spatial distributions.
However, observational bias may cause apparent diversity and lead to a wrong conclusion.
Although this study examines a single galaxy population, they are selected from the same dataset and method and free from observational bias; then, we have revealed that there is a wide range of diversity in the characteristics of protoclusters.
This will be an important hint to understand the physical nature of cluster formation.

\subsection{What controls Ly$\alpha$ strength?}
\begin{figure}
\includegraphics[width=\hsize]{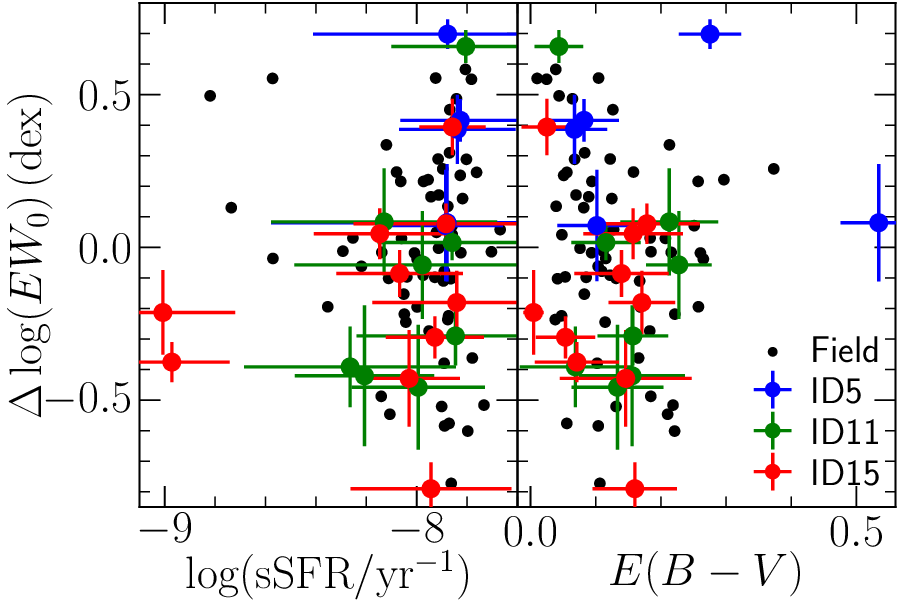}
\caption{Relation between Ly$\alpha$ strength, $\Delta\log(EW_0)$, and galaxy properties of star-forming activity
    (left) and dust attenuation (right).
    The ID5 protocluster (blue points) are biased toward higher $\Delta\log(EW_0)$, while the ID11 (green) and ID15
    (red) protoclusters are spread across lower $\Delta\log(EW_0)$, as presented in Figure~\ref{fig:EW}.
    However, there is no clear difference in the distributions of $sSFR$ and $E(B-V)$ between ID5 and the other
    protoclusters.}
\label{fig:EWgal}
\end{figure}

The strength of Ly$\alpha$ emissions are also significantly different among protoclusters as indicated in
Figure~\ref{fig:EW}.
In order to address the physical origin of these unique features, the deviation of Ly$\alpha$ $EW_0$ from the
average relation, $\Delta\log(EW_0)$, is compared with the galaxy properties of specific SFR, $sSFR=SFR/M_*$, and
dust extinction, $E(B-V)$ (Figure~\ref{fig:EWgal}).
As Ly$\alpha$ emission originate from ionising photons emitted by massive stars, star-forming activity is
one of the key physical conditions which control the strength of Ly$\alpha$ emissions.
On the other hand, Ly$\alpha$ emission tends to travel longer paths than UV radiation before escaping from a
galaxy due to resonant scattering, resulting in a higher sensitivity to dust attenuation for Ly$\alpha$ emission.
In contrast to these expectations, the $sSFR$ or $E(B-V)$ of the ID5 protocluster are not significantly different
from those of the ID11 and ID15 protoclusters (Figure~\ref{fig:EWgal}), while $\Delta\log(EW_0)$ clearly separates
the ID5 protocluster from the others (Figure~\ref{fig:EW}).
The ID5 protocluster tend to be distinguished from the ID15 protocluster by $sSFR$ though the statistical significance is not high ($p_\mathrm{KS}=0.061$).
Although there are many physical properties which cannot be estimated in this study (e.g., age, gas mass, or morphology),
no clear evidence is found that the unique nature of Ly$\alpha$ emission from the protocluster are attributable to the physical conditions of individual galaxies.

Inter-galactic medium can be another possible cause for the distinct Ly$\alpha$ properties seen in protocluster galaxies.
Even if Ly$\alpha$ emission is able to escape from their host galaxies, H{\sc i} gas confined in a protocluster
can dilute Ly$\alpha$ emissions.
\citet{shimakawa17} found only a few LAEs in a protocluster core at $z=2.53$ though many star-forming
galaxies traced by H$\alpha$ emissions populated the same region, suggesting that H{\sc i} gas depletes Ly$\alpha$ emissions.
\citet{cucciati14} more directly confirmed cold gas in a protocluster at $z=2.90$ by detecting Ly$\alpha$
absorption in the stacked spectrum of background galaxies.
Following these studies, cosmic structures are systematically mapped by utilising H{\sc i} absorption, resulting in the discoveries of protoclusters \citep[e.g.,][]{cai16,newman25}.
We cannot directly constrain the amount of H{\sc i} gas in a protocluster based on our available dataset.
Assuming that smaller $\Delta\log(EW_0)$ is caused by H{\sc i} gas, the member galaxies of the ID11 and ID15
protoclusters would be embedded in denser H{\sc i} gas.
The bottom panels of Figure~\ref{fig:galpropR} show the radial dependence of Ly$\alpha$ strength.
In the ID5 and ID11 protoclusters, member galaxies with lower $\Delta\log(EW_0)$ tend to be located in the inner
regions; particularly, the correlation between $\Delta\log(EW_0)$ and radial distance in the ID11 protocluster is
statistically significant according to Spearman's rank correlation test ($p=0.036$). This could be in line with a scenarios where H{\sc i} gas has thus far only accumulated in their core regions, resulting in a radial difference of H{\sc i}
gas absorption on Ly$\alpha$ emissions.
As for the ID15 protocluster, no correlation between $\Delta\log(EW_0)$ and radial distance is found, regardless of whether the geometric centre or the position of the most massive member is adopted as centre.
This may be interpreted as the ID15 protocluster having a large amount of H{\sc i} gas, so that almost all member
galaxies are embedded in the gas; thus, member galaxies in both core and outskirts are affected in a similar
manner irrespective of radial distance.
Therefore, it is possible to attribute the distinct Ly$\alpha$ properties among the three protoclusters
to a global feature of the protocluster rather than the physical properties within individual galaxies.
It should be noted that stronger Ly$\alpha$ emissions on the outskirts as seen in the ID5 or ID11 protoclusters would be consistent with \citet{cooke13} found.
The detailed investigation on specific protoclusters, like this study, is complementary to such a statistical study and necessary to identify the physical origin of such a phenomenon.

The variety of Ly$\alpha$ strength among the three protoclusters can be interpreted as a difference in
developmental phase: the ID5 and ID15 protoclusters are in a relatively earlier and later
developmental phases, respectively.
This interpretation seems to be consistent with what we have discussed based on star-forming activity of
protocluster galaxies (Section~\ref{sec:MS}).
A larger amount of H{\sc i} gas is naturally able to be linked to activating star formation; however, such tendency
of higher star-forming activity in the ID15 protocluster is not confirmed.
Since protoclusters are a complex of multiple galaxy populations, it is still possible that the ID15 protocluster
includes highly star-forming galaxies which cannot be probed by optical observations, like submm galaxies.

\section{Conclusions} \label{sec:con}
Following the construction of the systematic sample of protocluster candidates at $z\sim3\mathrm{-}5$ based on the
wide-field, deep imaging surveys of the HSC-SSP and CLAUDS, we have carried out follow-up spectroscopic
observations on three overdense regions of $U$-dropout ($z\sim3$) galaxies.
All three regions exhibit sharp peaks in their redshift distributions.
It is almost impossible to reproduce such redshift peaks by chance, indicating that they are physically associated.
Therefore, the three overdense regions of ID5, ID11, and ID15 are confirmed to be protoclusters composed of at
least six, nine, and ten member galaxies within the redshift range of $\Delta z=0.030$, 0.027, and 0.043,
respectively.
Since these three protoclusters are found from the same survey/dataset by identical techniques, they represent
good laboratories for studying the physical origin of the diversity of protocluster properties.
Our major results and implications are laid out below.

\begin{enumerate}
\item When all three protoclusters are combined into a single sample of protocluster members, there is no significant
    difference between protocluster and field galaxies neither in the distributions of $M_\mathrm{UV}$ nor
    Ly$\alpha$ $EW_0$.
    However, individual protoclusters exhibit their unique properties.
    The member galaxies of the ID5 protocluster have higher Ly$\alpha$ $EW_0$ at fixed $M_\mathrm{UV}$ compared
    with field galaxies while the ID11 and ID15 protoclusters tend to have lower Ly$\alpha$ $EW_0$.
    In contrast, no systematic difference in galaxy properties such as $sSFR$ or $E(B-V)$ are found among the protoclusters.
\item Among the three protoclusters, ID5 and ID11 host an AGN as a member, which is a higher fraction than that
    in the field.
    AGN activity may be more preferentially triggered in high-density environments due to e.g., galaxy mergers, or the AGN duty cycle may be enhanced by elevated gas accretion.
    Even between the ID5 and ID11 protoclusters, AGNs are differently placed near the centre or on the outskirts.
\item Deep, multi-wavelength imaging from $U$-band to IRAC/ch2 enables us to estimate physical properties (e.g.,
    $M_*$ and $SFR$) by SED fitting.
    Basically, most of protocluster galaxies follow the main sequence of star-forming galaxies as expected from
    the sample selection comprised of the Lyman break technique with follow-up Ly$\alpha$ spectroscopy. However, at the massive end,
    we can find a variety in galaxy properties among the three protoclusters: actively star-forming in ID5, but normal or suppressed
    in ID11 or ID15, respectively.
    Assuming that galaxies evolve from actively star-forming to quiescent as protoclusters mature, the
    ID5/ID15 protoclusters would be in an earlier/later developmental phase, respectively. 
\item The variation of the strength of Ly$\alpha$ emissions could be explained by the developmental phase of
    protoclusters.
    As the central halo of a protocluster grows by halo mergers, its deeper gravitational potential can hold a
    larger amount of H{\sc i} gas, like an intracluster medium.
    This global feature of protoclusters may result in coherent effects on their member galaxies' Ly$\alpha$ escape fractions.
    However, this is just one possible scenario, and further investigations are required to reach a firm conclusion.
\end{enumerate}

Galaxy clusters require on the order of a Hubble time to complete their formation as they are the most massive
objects in the Universe.
During such a long history of cluster formation, repeated halo mergers and mass accretion along the cosmic web occur in a complicated manner, so that even comparable mass clusters can follow different pathways of formation.
It is natural that galaxies residing in overdense regions also exhibit distinct features depending on the physical
conditions of their host protoclusters or surrounding cosmic web.
Although this study is based on only three protoclusters, our conclusions seem to be consistent with such a diverse context of
cluster formation.
In the future, sample size will be increased by e.g., Subaru/PFS and LSST, and we will be able to quantitatively
understand the physical processes of cluster formation. 

\section*{Acknowledgements}
We thank the anonymous referee for valuable comments and suggestions that significantly improved the manuscript.
JT acknowledges supports from Japan Society for the Promotion of Science (JSPS) KAKENHI grant numbers 25K07354 and Hyogo Science and Technology Association.
The Hyper Suprime-Cam (HSC) collaboration includes the astronomical communities of Japan and Taiwan, and Princeton University. The HSC instrumentation and software were developed by the National Astronomical Observatory of Japan (NAOJ), the Kavli Institute for the Physics and Mathematics of the Universe (Kavli IPMU), the University of Tokyo, the High Energy Accelerator Research Organization (KEK), the Academia Sinica Institute for Astronomy and Astrophysics in Taiwan (ASIAA), and Princeton University. Funding was contributed by the FIRST program from the Japanese Cabinet Office, the Ministry of Education, Culture, Sports, Science and Technology (MEXT), the Japan Society for the Promotion of Science (JSPS), Japan Science and Technology Agency (JST), the Toray Science Foundation, NAOJ, Kavli IPMU, KEK, ASIAA, and Princeton University.
This paper is based in part on data collected at the Subaru Telescope and retrieved from the HSC data archive system, which is operated by Subaru Telescope and Astronomy Data Center (ADC) at NAOJ. Data analysis was in part carried out with the cooperation of Center for Computational Astrophysics (CfCA) at NAOJ. We are honored and grateful for the opportunity of observing the Universe from Maunakea, which has the cultural, historical and natural significance in Hawaii.
This paper makes use of software developed for Vera C. Rubin Observatory. We thank the Rubin Observatory for making their code available as free software at \url{http://pipelines.lsst.io/}. 
The Pan-STARRS1 Surveys (PS1) and the PS1 public science archive have been made possible through contributions by the Institute for Astronomy, the University of Hawaii, the Pan-STARRS Project Office, the Max Planck Society and its participating institutes, the Max Planck Institute for Astronomy, Heidelberg, and the Max Planck Institute for Extraterrestrial Physics, Garching, The Johns Hopkins University, Durham University, the University of Edinburgh, the Queen’s University Belfast, the Harvard-Smithsonian Center for Astrophysics, the Las Cumbres Observatory Global Telescope Network Incorporated, the National Central University of Taiwan, the Space Telescope Science Institute, the National Aeronautics and Space Administration under grant No. NNX08AR22G issued through the Planetary Science Division of the NASA Science Mission Directorate, the National Science Foundation grant No. AST-1238877, the University of Maryland, Eotvos Lorand University (ELTE), the Los Alamos National Laboratory, and the Gordon and Betty Moore Foundation.
These data were obtained and processed as part of the CFHT Large Area $U$-band Deep Survey (CLAUDS), which is a collaboration between astronomers from Canada, France, and China described in \citet{sawicki19}. CLAUDS is based on observations obtained with MegaPrime/MegaCam, a joint project of CFHT and CEA/DAPNIA, at the CFHT which is operated by the National Research Council (NRC) of Canada, the Institut National des Science de l'Univers of the Centre National de la Recherche Scientifique (CNRS) of France, and the University of Hawaii. CLAUDS uses data obtained in part through the Telescope Access Program (TAP), which has been funded by the National Astronomical Observatories, Chinese Academy of Sciences, and the Special Fund for Astronomy from the Ministry of Finance of China. CLAUDS uses data products from TERAPIX and the Canadian Astronomy Data Centre (CADC) and was carried out using resources from Compute Canada and Canadian Advanced Network For Astrophysical Research (CANFAR).

\section*{Data Availability}
The data underlying this article will be shared on reasonable request to the corresponding author.

\bibliographystyle{mnras}
\bibliography{refs}

% Don't change these lines
\bsp	% typesetting comment
\label{lastpage}
\end{document}